\documentclass[a4paper,11pt,twoside]{article}
\usepackage[paperheight=29.7cm,paperwidth=21cm,outer=2.25cm,inner=2.25cm,top=2cm,bottom=2cm,headheight=13.6pt,includehead,includefoot]{geometry} 
\usepackage{silence}
\newcommand{\textlineskip}{\baselineskip=13pt}

\providecommand{\keywords}[1]{\noindent\textbf{Keywords:} #1}

\def\fnt#1#2{\footnotetext{\kern-.3em%
          {$^{\mbox{\scriptsize #1}}$}{#2}}}
          
\newcommand{\fcaption}[1]{\caption{#1}}
\newcommand{\tcaption}[1]{\caption{#1}}

\usepackage{fancyhdr}

\fancypagestyle{pagestyle}{%
  \fancyhf{}
  \fancyhead[OR]{E. Tolotti, E. Zardini, E. Blanzieri, D. Pastorello}
  \fancyhead[EL]{Ensembles of Quantum Classifiers}
  \fancyfoot[C]{\thepage}%
}

\fancypagestyle{firstpagestyle}{%
  \fancyhf{}
  
  \fancyfoot[C]{\thepage}
}

\renewcommand{\thefootnote}{\fnsymbol{footnote}}  

\usepackage{amsmath}
\usepackage{amssymb}
\usepackage[linesnumbered]{algorithm2e}
\usepackage[font={footnotesize}]{caption}

\usepackage{bm}

\usepackage{multirow}
\usepackage[figuresright]{rotating}

\usepackage[hyphens]{url}
\usepackage[hyperfootnotes=false]{hyperref}
\usepackage{xcolor}
\hypersetup{
  colorlinks,
  linkcolor={black},
  citecolor={black},
  urlcolor={black}
}
\usepackage[noabbrev,capitalise]{cleveref}

\usepackage{soul}

\usepackage{etoolbox}
\usepackage{accents}
\robustify{\underaccent}

\usepackage{braket}

\usepackage{tikz}
\usetikzlibrary{quantikz2}

\usepackage[caption=false]{subfig}

\begin{document}

\normalsize\textlineskip
\thispagestyle{firstpagestyle}
\setcounter{page}{1}

\centerline{\Large\bf
Ensembles of Quantum Classifiers}

\vspace{25pt}

\begin{center}
    \textbf{Emiliano Tolotti$^{\star \|}$}, \hspace{1pt}
    \textbf{Enrico Zardini$^{\star}$}, \\
    \textbf{Enrico Blanzieri$^{\star \dagger}$}, \hspace{1pt}
    \textbf{Davide Pastorello$^{\dagger \ddagger}$}

    \vspace*{10pt}

    $^\star$ Department of Information Engineering and Computer Science\\ University of Trento \\ 
    $ $ via Sommarive 9, 38123 Povo, Trento, Italy

    \vspace*{10pt}

    $^\dagger$ Trento Institute for Fundamental Physics and Applications \\ 
    $ $ via Sommarive 14, 38123 Povo, Trento, Italy

    \vspace*{10pt}

    $^\ddagger$ Alma Mater Studiorum - Università di Bologna \\ 
    $ $ Piazza di Porta San Donato 5, 40126 Bologna, Italy

    \vspace*{10pt}

    $^\|$ emiliano.tolotti@unitn.it
\end{center}

\vspace*{15pt}

\begin{abstract}
\noindent
In the current era, known as Noisy Intermediate-Scale Quantum (NISQ), encoding large amounts of data in the quantum devices is challenging and the impact of noise significantly affects the quality of the obtained results. A viable approach for the execution of quantum classification algorithms is the introduction of a well-known machine learning paradigm, namely, the ensemble methods. Indeed, the ensembles combine multiple internal classifiers, which are characterized by compact sizes due to the smaller data subsets used for training, to achieve more accurate and robust prediction performance. In this way, it is possible to reduce the qubits requirements with respect to a single larger classifier while achieving comparable or improved performance. In this work, we present an implementation and an extensive empirical evaluation of ensembles of quantum classifiers for binary classification, with the purpose of providing insights into their effectiveness, limitations, and potential for enhancing the performance of basic quantum models. In particular, three classical ensemble methods and three quantum classifiers have been taken into account here. Hence, the scheme that has been implemented (in Python) has a hybrid nature. The results (obtained on real-world datasets) have shown an accuracy advantage for the ensemble techniques with respect to the single quantum classifiers, and also an improvement in robustness. In fact, the ensembles have turned out to be able to mitigate both unsuitable data normalizations and repeated measurement inaccuracies, making quantum classifiers more stable.

\vspace*{10pt}
\end{abstract}

\setcounter{footnote}{0}
\renewcommand{\thefootnote}{\alph{footnote}}

\vspace*{1pt}\textlineskip    

\section{Introduction}
\label{sec:introduction}
Quantum machine learning (QML) is a recent field of research, which aims at developing quantum algorithms for solving machine learning problems in a more efficient way than the classical counterparts \cite{Biamonte_2017}. If a sufficient number of fully connected qubits were available, a quantum advantage (with respect to classical supercomputers) could be achieved on different tasks and practical problems could be tackled effectively. However, the current era, known as Noisy Intermediate-Scale Quantum (NISQ) \cite{Preskill_2018}, is characterised by noisy devices with limited numbers of qubits. Additionally, in gate-based quantum devices, the impact of noise increases with the circuit's depth and size. As a consequence, encoding large amounts of data turns out to be challenging, and the quality of the results obtained is significantly affected by noise. To execute quantum algorithms on the current architectures, the size of the circuits must be reduced. In this way, despite the current limitations, practical problems could be addressed, making progress towards solving real-world challenges using quantum computing technologies. 

Ensemble methods are a widely used machine learning technique that consists in combining the predictions of multiple models \cite{1688199}. This approach aims at enhancing prediction accuracy and stability by exploiting model diversity. In the context of quantum computation, classical ensembles represent an effective way to reduce the computational requirements. Indeed, the internal models are typically characterized by compact sizes due to the smaller data subsets used for training. In practice, ensemble methods allow executing quantum algorithms on NISQ devices thanks to a more efficient resource usage with respect to single larger models (the quantum circuits involved are smaller). 

Quantum ensemble methods have been also developed. For instance, Schuld and Petruccione \cite{Schuld2018} and Abbas et al. \cite{Abbas2020} have proposed quantum ensemble classifiers based on Bayesian Model Averaging (BMA). The ensembles in question exploit non-trainable classifiers, under the assumption that a large ensemble of weak classifiers can achieve good performance. Instead, Araujo and da Silva have presented a quantum ensemble of trainable classifiers \cite{araujo2020quantum}. In particular, they have considered a superposition of quantum classifiers, with the classifiers being quantum neural networks. Macalauso et al. have proposed a quantum ensemble framework \cite{macaluso2022quantum} that is based on bagging and is characterised by an exponential growth of the ensemble size at the price of a linear increase in the circuit depth. Regarding the work by Windridge and Nagarajan \cite{10.1007/978-3-319-52289-0_9}, a quantum-SVM-based attribute bootstrap aggregation is presented. In practice, a superposition of quantum decision hyperplanes is used to perform attribute selection. Eventually, Qin et al. \cite{qin2022improving} and Zhang and Wang \cite{zhang2022efficient} have proposed hybrid techniques to efficiently combine quantum classification algorithms, showing how the parallel combination of multiple variational quantum classifiers can outperform state-of-the-art classification methods.

In this work, we present an implementation and the empirical evaluation of ensembles of quantum classifiers. In detail, the proposed scheme is hybrid, with classical ensemble methods and quantum classification algorithms. Regarding the ensemble methods, bootstrap \cite{bootstrap}, boosting \cite{schapire2013explaining}, and stacking \cite{stacking} have been taken into account. Concerning the quantum classifiers, a quantum cosine classifier \cite{Pastorello_2021}, a quantum distance classifier \cite{Schuld_2017}, and a quantum $k$-nearest neighbors (quantum $k$-NN) classifier \cite{Afham2020, Ma2021} have been considered. The quantum algorithms employed in this work, like many other quantum machine learning algorithms, require the existence of a quantum random access memory (QRAM) \cite{Giovannetti_2008} in order to achieve a speedup with respect to classical computation. Working prototypes of QRAMs have not been developed yet. However, this work does not focus on the potential quantum time advantage, but on the empirical evaluation of the considered scheme in terms of accuracy. To this end, the methods have been evaluated on a binary classification task on real-world datasets. The results have shown an accuracy advantage with respect to single classifiers and also an improvement in robustness. Indeed, the ensembles have shown the capability of mitigating both unsuitable data normalizations and repeated measurement inaccuracies.

The article is structured as follows: \cref{sec:background} provides some background information; \cref{sec:ensables-of-classifiers} introduces the hybrid scheme and presents the implementation details; \cref{sec:empirical-evaluation} deals with the experiments performed and the results obtained; \cref{sec:conclusions} concludes the work.

\section{Background}
\label{sec:background}
In this section, some background information about quantum information and quantum machine learning is provided. Then, the algorithms considered in this work, which include classical ensemble methods and quantum classifiers, are introduced.

\subsection{Quantum machine learning}
\label{subsec:qml}
Quantum computing is a type of computation that exploits the principles of quantum mechanics, such as superposition and entanglement, to perform calculations. In 2013, Lloyd, Mohseni, and Rebentrost showed that quantum computing can be used to obtain an exponential speedup with respect to classical clustering algorithms \cite{https://doi.org/10.48550/arxiv.1307.0411}. This sparked the interest in the usage of quantum computers to enhance machine learning and marked the emergence of QML, which is now a key area of research.

Roughly speaking, in quantum information, the qubit is the basic unit of information, analogously to the bit for classical information. More precisely, a qubit is any quantum system that can be described in a $2$-dimensional Hilbert space. The quantum states are in bijective correspondence with the projective rays in the Hilbert space. Hence, the state of a qubit can be represented by a unit vector in the 2-dimensional complex vector space $\mathbb{C}^2$, whose standard orthonormal basis vectors, which are denoted as $\ket{0}$ and $\ket{1}$, form the so-called computational basis. As a consequence, a qubit can be not only in one of the two basis states, but also in any superposition, i.e., linear combination, of them. In the Dirac notation, this is written as
\begin{equation*}
    \ket{\psi} = \alpha \ket{0} + \beta \ket{1},
\end{equation*}
where $\alpha, \beta \in \mathbb{C}$ are called amplitudes and must satisfy the constraint $|\alpha|^2 + |\beta|^2 = 1$. Actually, the amplitudes values cannot be directly observed. In fact, when a qubit is measured, one of the basis states ($\ket{0}$ or $\ket{1}$) is obtained, with probabilities $|\alpha|^2$ and $|\beta|^2$, respectively. Hence, they can only be estimated by performing repeated measurements. The computational advantages emerge when considering composite systems featuring multiple qubits. Indeed, a system of $n$ qubits, also known as \emph{register}, is described in the tensor product Hilbert space $(\mathbb C^2)^{\otimes n}$. Therefore, its state can be described by a vector of size $2^n$:
\begin{equation}\label{eq:product}
    \bigotimes_{i=0}^{n-1} \ket{\psi_i} = \ket{\psi_0...\psi_{n-1}}\qquad \ket{\psi_i}\in \mathbb C^2\quad\forall i=0,\dots,n-1.
\end{equation}
However, not all the unit vectors in $(\mathbb C^2)^{\otimes n}$ can be decomposed into the product form (\ref{eq:product}). The states of the $n$-qubit register described by non-product vectors are called \emph{entangled states} and encode non-classical correlations among the qubits. 
In practice, by operating on quantum registers characterised by state superposition, it is possible to perform parallel operations, and the presence of quantum entanglement is the key enabling quantum advantages with respect to classical computations. 

In quantum circuits, which represent the most common quantum computation model, the computation is performed by means of quantum gates, which implement unitary operations. An important one-qubit gate is the \textit{Hadamard} gate, which creates a balanced superposition of the computational basis states and is defined as
\begin{equation*}
    H = \frac{1}{\sqrt{2}} \begin{bmatrix} 1 & 1 \\ 1 & -1 \end{bmatrix}.
\end{equation*}
Instead, a relevant two-qubit gate is the controlled-\textit{NOT} gate (\textit{CNOT}), which flips the state of the target qubit if the control qubit is in state $\ket{1}$. It is defined as
\begin{equation*}
    CNOT = \begin{bmatrix} 1 & 0 & 0 & 0\\ 0 & 1 & 0 & 0\\ 0 & 0 & 0 & 1\\ 0 & 0 & 1 & 0\end{bmatrix}
\end{equation*}
and, when combined with the Hadamard gate, it can create entanglement.  
In particular, there exist sets of quantum gates that are universal, in the sense that any $n$-qubit quantum gate can be implemented up to arbitrary precision by using gates taken from these sets. For example, the set \textit{\{H, $P_\phi$, CNOT\}}, where $P_\phi$ is the phase shift gate, is universal \cite{nielsen_chuang_2010}.

A non-trivial problem in quantum computing (still unsolved) is how to encode data into quantum states. In particular, there are two main strategies: basis encoding and amplitude encoding. In basis encoding, data is encoded into the computational basis as classical bits. Hence, given a binary string $x = (b_1,...,b_n)$, with $b_i \in \{0, 1\}$, $x$ is encoded as $\ket{x} = \bigotimes_{i=1}^n\ket{b_i}$. In addition, by exploiting superposition, it is possible to represent a dataset $X = \{x_1,...,x_m\}$ as $\ket{X} = \frac{1}{\sqrt{m}}\sum_{i=1}^m\ket{x_i}$. Instead, in amplitude encoding, data is stored into the amplitudes of quantum states. Given a normalized data vector $x$, i.e., a vector $x \in \mathbb{C}^d$ such that
$||x|| = 1$, it is encoded as $\ket{x} = \sum_{i=1}^d x_i \ket{i}$.
On the one hand, within the amplitude encoding framework, $d$-dimensional data vectors can be encoded in $log(d)$ qubits. On the other hand, amplitudes are not directly observable; therefore, a repeated sampling of the qubits state is necessary in order to estimate the amplitudes values.

As stated previously, in QML, quantum computing procedures are exploited to improve machine learning algorithms. An important procedure is the so-called SWAP test \cite{PhysRevLett.87.167902}, which allows estimating the dot product of two data vectors. Specifically, the corresponding quantum circuit, which is also the basis of two quantum classifiers employed in this work, is the following: 
\begin{figure}[h!]
    \centering
    \begin{quantikz}
    \lstick{$\ket{0}$} & \gate{H} & \ctrl{2} &  \gate{H} & \meter{} \qw \\
    \lstick{$\ket{\psi}$} & \qw & \swap{1} & \qw & \qw \\
    \lstick{$\ket{\phi}$} & \qw & \targX{} & \qw & \qw
    \end{quantikz}
\end{figure}

\noindent Given the quantum states $\ket{\psi}$ and $\ket{\phi}$ (that can be also $n$-qubit states), the probability of measuring 0 on the ancillary qubit after performing the SWAP test is $\mathbb{P}(0) = \frac{1}{2} + \frac{1}{2}|\braket{\psi|\phi}|^2$. Hence, by executing the circuit multiple times, it is possible to estimate the squared inner product of the quantum states, which corresponds to the squared dot product of the data vectors encoded in the amplitudes of $\ket{\psi}$ and $\ket{\phi}$. In detail, in order to obtain an estimate up to an error $\epsilon$, the number of circuit repetitions required is $O(\frac{1}{\epsilon^2})$.

Eventually, many QML algorithms, including one classifier used in this work, assume the existence of a quantum random access memory \cite{Giovannetti_2008} for an efficient state preparation. The idea is to query a superposition of addresses to retrieve a superposition of $N=2^n$ memory cells in time $O(n)$. Some physical proposals have been suggested, but there is no working implementation yet and there are still doubts on its actual feasibility.

\subsection{Ensemble techniques}
\label{subsec:ensemble-techniques}
Ensemble learning is a machine learning paradigm based on the intuition that combining multiple models is more effective than using a single model \cite{1688199}. Indeed, weak base models can suffer from high bias or high variance, but their combination can produce a strong and more robust learner with good performance. The ensemble techniques taken into account in this work are bootstrap, boosting and stacking.

\subsubsection{Bootstrap aggregating} 
\label{subsubsec:bootstrap}
The bootstrap aggregating algorithm, also known as bagging, is a simple ensemble scheme based on the bootstrap sampling procedure proposed by Breiman \cite{bootstrap}. In practice, homogeneous internal models are independently trained on sets obtained from the training set by random sampling with replacement. This means that an element can be sampled multiple times, and subsequent samplings are independent (i.i.d. samples). For a regression task, the predicted value is the average of the models outputs. Instead, for a classification task, a majority voting scheme is used. In particular, in the case of a binary classification task with labels in $\{-1, +1\}$, the majority voting can be expressed as
\begin{equation}
    y(x) = sign\left(\sum_{i = 1}^M m_i(x)\right),
    \label{eq:bootstrap-pred}
\end{equation}
where $M$ is the number of internal models, and $m_i$ is the $i$-th internal model. By introducing diversity in the data through the bootstrapping process, bagging decreases the variance, improving the performance and the robustness. In addition, it is a scalable and parallelizable algorithm. Indeed, both the ensemble building (the sampling of the training instances for the classifiers is independent) and the prediction step can be parallelized. 

\subsubsection{Boosting} 
\label{subsubsec:boosting}
Boosting is a homogeneous ensemble model based on an iterative training procedure. In detail, at each iteration, a weak classifier is trained on data sampled from the training set according to a distribution that is influenced by the performance of the previous iteration classifiers. The boosting algorithm used in this work is one of the most famous ones, namely, AdaBoost \cite{schapire2013explaining}. In AdaBoost, which stands for adaptive boosting, a weak classifier is trained on a subset of the training set and used to predict the class (let us restrict to classification) of all training samples. The classification errors are then used to increase the weight of the misclassified instances for the next iteration sampling and to compute a weight for the classifier inside the final aggregated model. Indeed, the aggregation strategy is a weighted average of the internal models. In particular, for a binary classification task with labels in $\{-1, +1\}$, the ensemble prediction can be written as
\begin{equation}
    y(x) = sign\left(\sum_{i = 1}^M \alpha_i m_i(x)\right),
    \label{eq:boosting-pred}
\end{equation}
where $\alpha_i$ is the weight of the $i$-th internal model $m_i$. Boosting decreases the bias, while also reducing the variance. Nevertheless, the training procedure, which is iterative, cannot be parallelized. Instead, at prediction time, the output of the internal classifiers can be computed in parallel. 

\subsubsection{Stacking} 
\label{subsubsec:stacking}
Stacking is a heterogeneous ensemble technique \cite{stacking}. In particular, different internal models are trained and evaluated on the training set using a $k$-fold cross validation technique, obtaining a prediction for every internal classifier - training point pair. These predictions are then used as the training set of a meta-classifier that combines the output of the internal classifiers into a final prediction. Specifically, the internal classifiers are trained on the full training set, and the stacking classifier prediction can be expressed as
\begin{equation}
    y(x) = m_{meta}(m_1(x),...,m_M(x)),
    \label{eq:stacking-pred}
\end{equation}
where $m_{meta}$ is the meta-classifier model. The main advantage of stacking is that, by combining diverse classifiers based on different assumptions, the performance with respect to the single classifiers improve. In addition, the internal classifiers can be trained and executed in parallel.

\subsection{Quantum classifiers}
\label{subsec:quantum-classifiers}
The quantum classification algorithms considered are a quantum cosine classifier, a quantum distance classifier, and a quantum $k$-nearest neighbors classifier. Their details are provided below. 

\subsubsection{Quantum cosine classifier} 
\label{subsubsec:q-cosine-classifier}
The quantum cosine classifier proposed by Pastorello and Blanzieri \cite{Pastorello_2021} is an algorithm for binary classification based on the cosine similarity of data vectors. In particular, the classification function implemented by the classifier is the following:
\begin{equation}
    y(x) = sign \left(\sum_{i=0}^{N-1}{y_i \cos{(x_i, x)}}\right),
    \label{eq:cos-class-classical-formula}
\end{equation}
where $N$ is the number of training samples, $x_i$ is the feature vector of the $i$-th sample, $y_i \in \{-1, +1\}$ is the corresponding label, and
\begin{equation*}
\cos{(x, y)} = \frac{x \cdot y}{\lVert x\rVert \lVert y\rVert}.
\end{equation*}

Concerning the data encoding scheme, the classifier uses the amplitude encoding for the feature vectors $x_i$, which must be unit-norm normalized, and the basis encoding for the binary labels $y_i$, which are mapped to the domain $\{1, 0\}$ according to
\begin{equation}
    l_i = \frac{1 - y_i}{2}.
    \label{eq:label-to-qubit-state}
\end{equation}
The initial state is defined as
\begin{equation*}
    \ket{\Psi} = \frac{1}{\sqrt{2}} (\ket{\psi_x}\ket{0} + \ket{\psi}\ket{1}) \in \mathcal{H}_n \otimes \mathcal{H}_d \otimes \mathcal{H}_l \otimes \mathcal{H}_a ,
\end{equation*}
where 
\begin{align*}
    \ket{\psi_x} &= \frac{1}{\sqrt{N}} \sum_{i=0}^{N-1}{\ket{i}\ket{x_i}\ket{l_i}} \in \mathcal{H}_n \otimes \mathcal{H}_d \otimes \mathcal{H}_l,\,\textnormal{\ and} \\
    \ket{\psi} &= \frac{1}{\sqrt{N}} \sum_{i=0}^{N-1}{\ket{i} \ket{x} \ket{-}} \in \mathcal{H}_n \otimes \mathcal{H}_d \otimes \mathcal{H}_l .
\end{align*}
In detail, the classifier assumes the existence of a QRAM in order to have an efficient preparation of the initial state. Then, a SWAP test on the states $\ket{+}$ and $\ket{\Psi}$ is performed, and the probability of measuring $1$ on the SWAP test ancillary qubit turns out to be
\begin{equation*}
    \mathbb{P}(1) = \frac{1}{4}(1 - \braket{\psi_x|\psi}),
\end{equation*}
with 
\begin{equation*}
    \braket{\psi_x|\psi} = \frac{1}{\sqrt{2}N}\sum_{i=0}^{N-1}{l_i\cos{(x_i, x)}}.
\end{equation*}
Eventually, the predicted label is given by
\begin{equation*}
    y(x) = sign(1 - 4\mathbb{P}(1)).
\end{equation*}

In practice, the quantum circuit requires $3$ qubits for the SWAP test, $\left\lceil \log_2{N} \right\rceil$ qubits for the index register ($\mathcal{H}_n$), $\left\lceil \log_2{D} \right\rceil$ qubits for the feature register ($\mathcal{H}_d$, with $D$ being the number of features) and $1$ qubit for the binary labels ($\mathcal{H}_l$). If a QRAM is available, the algorithm has a time complexity of $O(\log(ND))$ and a space complexity of $O(\log(ND))$ qubits.

\subsubsection{Quantum distance classifier} 
\label{subsubsec:q-distance-classifier}
The quantum distance classifier proposed by Schuld et al. \cite{Schuld_2017} is a binary classification algorithm based on the squared euclidean distance of feature vectors. In this work, a slightly modified version is considered. Specifically, the classification function implemented by the classifier is defined as
\begin{equation}
    y(x) = sign \left(\sum_{i=0}^{N-1}{y_i(1 - \frac{1}{4}||x_i - x||^2})\right),
    \label{eq:dist-class-classical-formula}
\end{equation}
where $N$ is the number of training samples, $x_i$ is the feature vector of the $i$-th sample, and $y_i \in \{-1, +1\}$ is the corresponding label. 

As in the quantum cosine classifier, the amplitude encoding is used for the features vector, the basis encoding is used for the binary labels (the mapping is given by Eq. (\ref{eq:label-to-qubit-state})), and the presence of a QRAM is assumed. The initial state of the circuit is defined as
\begin{equation*}
    \ket{\Psi} = \frac{1}{\sqrt{2N}} \sum_{i=0}^{N-1} \ket{i}(\ket{0}\ket{x} + \ket{1}\ket{x_i})\ket{l_i} \in \mathcal{H}_n \otimes \mathcal{H}_a \otimes \mathcal{H}_d \otimes \mathcal{H}_l.
\end{equation*}
In particular, the quantum circuit consists of a Hadamard gate and two qubits measurements, with the second one being a conditional measurement; thus, the circuit complexity is constant. The probability of obtaining $k \in \{0, 1\}$ in the second measurement, which is performed on the label qubit, is equal to
\begin{equation*}
    \mathbb{P}(k) = \frac{1}{4Np_0} \sum_{i:l_i=k} ||x + x_i||^2,
\end{equation*}
where $p_0$ is the probability of obtaining $0$ in the first measurement. Since the data vectors are characterised by unit norm, the following relationship holds:
\begin{equation*}
    \frac{1}{4N} \sum_{i} ||x + x_i||^2 = \frac{1}{N} \sum_{i} (1 - \frac{1}{4}||x - x_i||^2).
\end{equation*}
Hence, the predicted label is given by
\begin{equation*}
    y(x) = sign\left(\mathbb{P}(0|0) - \frac{1}{2}\right).
\end{equation*}

Basically, the quantum circuit needs $1$ ancillary qubit ($\mathcal{H}_a$), $\left\lceil \log_2{N} \right\rceil$ qubits for the index register ($\mathcal{H}_n$), $\left\lceil \log_2{D} \right\rceil$ qubits for the feature register ($\mathcal{H}_d$, with $D$ being the number of features), and $1$ qubit to encode the binary labels ($\mathcal{H}_l$). Assuming the availability of a QRAM, the algorithm has a time complexity of $O(\log(ND))$ and a space complexity of $O(\log(ND))$ qubits.

\subsubsection{Quantum k-nearest neighbors classifier} 
\label{subsubsec:q-k-nn-classifier}
The quantum $k$-nearest neighbors classifier proposed by Afham et al. \cite{Afham2020} and Ma et al. \cite{Ma2021} is a quantum version of the $k$-nearest neighbors algorithm \cite{1053964}, which is one of the simplest algorithms for multiclass classification in machine learning. In particular, the quantum $k$-NN in question is based on the notion of fidelity of quantum states. Indeed, the algorithm selects the $k$ nearest neighbors based on the fidelity of the states encoding the training and test feature vectors, with the fidelity $F$ being defined as
\begin{equation*}
    F(\ket{\psi}, \ket{\phi}) = |\braket{\psi|\phi}|^2.
\end{equation*}

More in detail, the algorithm exploits the amplitude encoding for the data features, and the initial state is defined as 
\begin{equation*}
    \ket{\Psi} = \ket{0}\ket{x}\ket{\psi_x} \in \mathcal{H}_a \otimes \mathcal{H}_d \otimes \mathcal{H}_d \otimes \mathcal{H}_n,
\end{equation*}
where
\begin{equation*}
    \ket{\psi_x} = \frac{1}{\sqrt{N}} \sum_{i=0}^{N-1} \ket{x_i}\ket{i} \in \mathcal{H}_d \otimes \mathcal{H}_n,
\end{equation*}
$N$ is the number of training samples, and $x_i$ is the feature vector of the $i$-th sample. Regarding the quantum circuit, it consists of a SWAP test and two measurements. Specifically, the first measurement is performed on the SWAP test ancillary qubit ($\mathcal{H}_a$), while the second one is performed on the index register $\ket{i}$. By iterating this procedure, it is possible to estimate, for each index $i$, the quantity $\mathbb{Q}(i)$, which is defined as
\begin{equation*}
    \mathbb{Q}(i) = \mathbb{P}(i|0) - \mathbb{P}(i|1) = \frac{2(F_i - \braket{F})}{N(1 - \braket{F}^2)},
\end{equation*}
where $F_i$ is the fidelity of the quantum states $\ket{x}$ and $\ket{x_i}$, and $\braket{F}$ is the average value of $F_i$ over $i$. Eventually, the $k$ nearest neighbors are retrieved by classically sorting the training data according to $\mathbb{Q}(i)$, and the predicted label is obtained through a majority voting.

In practice, the circuit requires $1$ ancillary qubit for the SWAP test ($\mathcal{H}_a$), $\left\lceil \log_2{N} \right\rceil$ qubits for the index register ($\mathcal{H}_n$), and $2\left\lceil \log_2{D} \right\rceil$ qubits for the feature registers ($\mathcal{H}_d$, with $D$ being the number of features). Assuming the presence of a QRAM, the circuit time complexity is $O(\log(ND))$ and the space complexity is $O(\log(ND))$ qubits.

\section{Ensembles of Quantum Classifiers}
\label{sec:ensables-of-classifiers}
\begin{figure}[htb]
    \centering
    \includegraphics[width=0.95\linewidth]{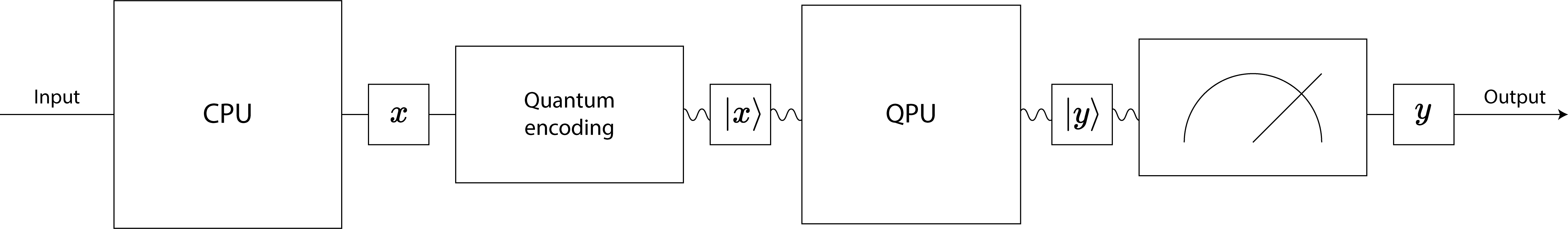}
    \vspace*{10pt}
    \fcaption{High-level view of the interaction between classical and quantum components in the proposed hybrid scheme.}
    \label{fig:hybrid-cl-q-interaction}
\end{figure}

In this work, a hybrid scheme characterised by classical ensembles and quantum internal classifiers is introduced. In practice, the quantum classifiers described in \cref{subsec:quantum-classifiers} are used as internal models of the ensemble methods presented in \cref{subsec:ensemble-techniques}. A high-level view of the interaction between classical and quantum components is shown in \cref{fig:hybrid-cl-q-interaction}. First of all, the input data is classically processed to produce the input for the quantum models. Then, after the quantum encoding of the classical information, the quantum circuits are run multiple times, with final measurements, in order to obtain sufficiently precise estimates of the output quantities. Eventually, the output of the quantum models is classically post-processed to either carry on the training procedure or provide the final output.

The hybrid scheme allows employing quantum classification algorithms in homogeneous and heterogeneous ensemble techniques taken from the literature. In this way, it is possible to analyse the advantages in accuracy and robustness of using quantum classifiers in ensemble schemes while being compatible with the hardware limitations of current architectures, which do not allow efficient quantum implementations of ensemble techniques yet.

\subsection{Implementation}
\label{subsec:implementation}
The hybrid scheme has been developed in Python language using Qiskit \cite{Qiskit}, the open-source SDK provided by IBM for building and running quantum circuits either on quantum hardware \cite{ibmquantum} or in simulation (the code is available at \url{https://github.com/emiliantolo/ensembles-quantum-classifiers}). In this work, the high-performance Aer simulator has been used for the execution of the algorithms. It is also worth mentioning that, although the scheme is theoretically valid for multiclass classification tasks, the code provided here supports only binary classification. Additional details about the implementation of the models are provided in the following. 

\begin{figure}[p!]
    \centering
    \subfloat[\label{fig:q-cos-class-circ}]{
        \centering
        \includegraphics[width=0.85\linewidth]{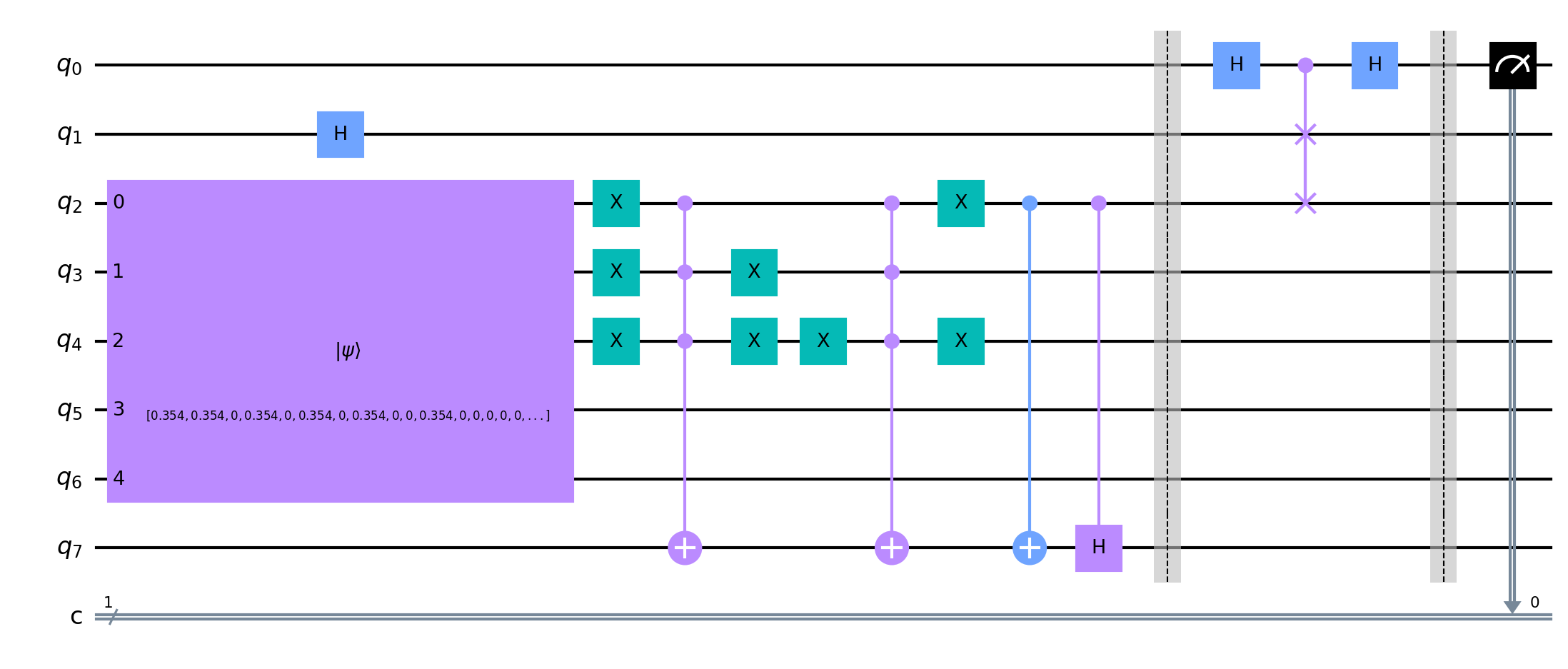}
    } \\ 
    \subfloat[\label{fig:q-dist-class-circ}]{
        \centering
        \includegraphics[width=0.85\linewidth]{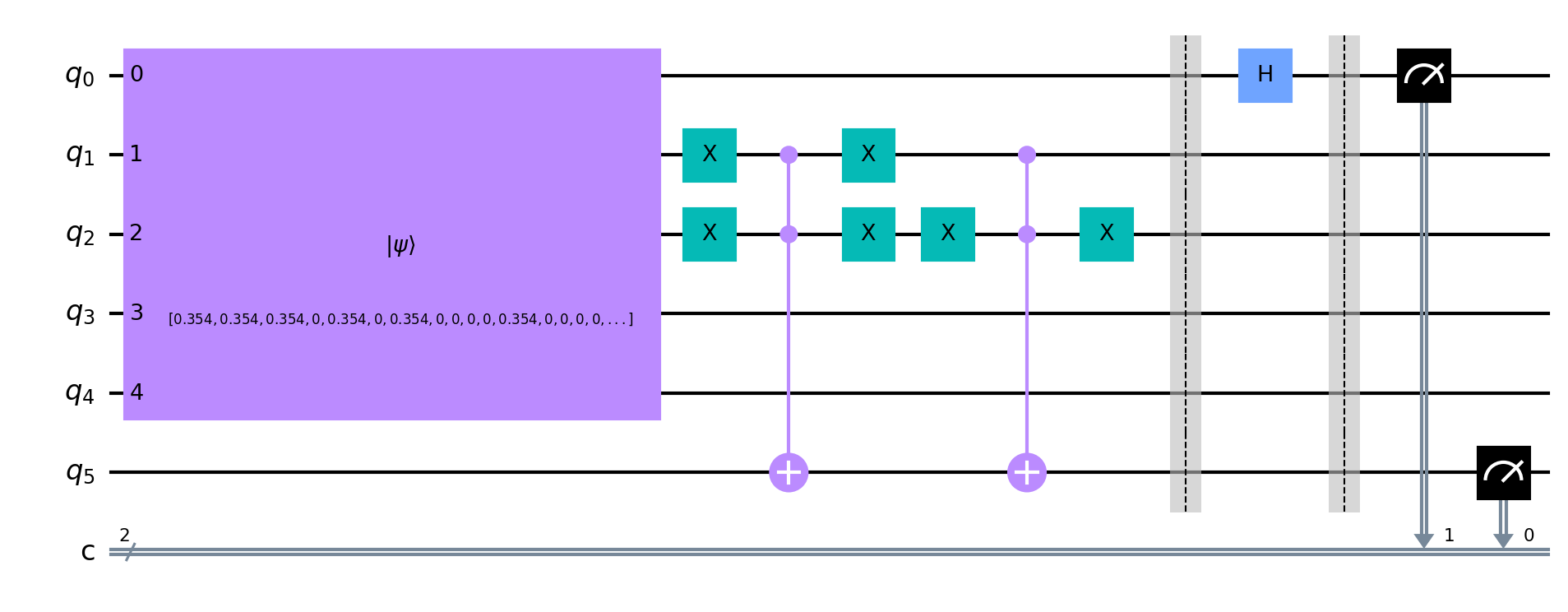}
    } \\ 
    \subfloat[\label{fig:q-k-nn-class-circ}]{
        \centering
        \includegraphics[width=0.85\linewidth]{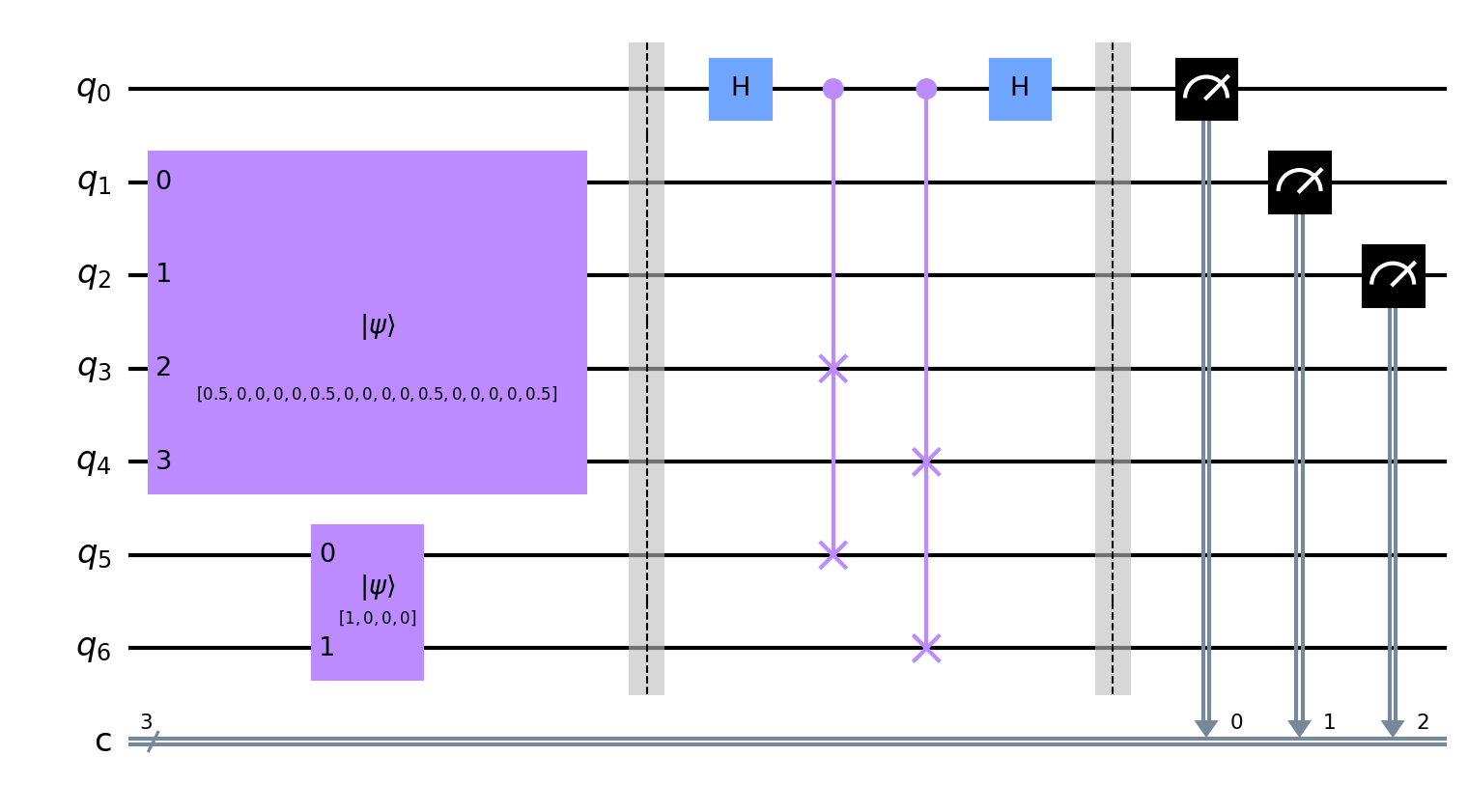}
    } \vspace{4pt}
    \fcaption{Quantum circuit example for the quantum cosine classifier (a), the quantum distance classifier (b), and the quantum $k$-NN classifier; the dataset and the test instance considered are $X = \{([1, 0, 0, 0], -1), ([0, 1, 0, 0], -1), ([0, 0, 1, 0], 1), ([0, 0, 0, 1], 1)\}$ and $x = [1,0,0,0]$, respectively.}
    \label{fig:q-classifiers-circs}
\end{figure}

\subsubsection{Ensemble techniques}
\label{subsubsec:ensembles-implementation}
Ensemble methods have been developed form scratch, ensuring a standard implementation. Regarding bootstrap (\cref{subsubsec:bootstrap}), the ensemble is built by sampling with replacement $N$ subsets of $S$ elements each from the training set; these subsets are then used as the training sets of the $N$ internal quantum classifiers. Concerning boosting (\cref{subsubsec:boosting}), $N$ training iterations are performed. In detail, at each step, a classifier is trained on a subset of $S$ elements sampled from the training set according to the distribution determined by the previous iterations; then, the classifier is used to predict the training set labels, allowing the computation of the classifier weight in the ensemble and the definition of the new distribution. It is worth mentioning that an $\epsilon = 1e-10$ value has been added in the computation of errors, in order to avoid divisions by zero. Eventually, for stacking, a $k$-fold cross validation procedure is run on the training set for each selected internal model, obtaining a prediction for each training instance; these predictions are then used as the training set for the meta-classifier, while the internal models are trained on the full training set. In particular, the meta-model used in this work takes as input not only the predicted output classes, but also the prediction confidences of the internal models. Concerning the prediction step, it is performed according to Eqs. \eqref{eq:bootstrap-pred}, \eqref{eq:boosting-pred}, and a variation of \eqref{eq:stacking-pred}, respectively.

\subsubsection{Quantum classifiers}
\label{subsubsec:quantum-classifiers-implementation}
First of all, the implementation of the quantum cosine classifier described in \cref{subsubsec:q-cosine-classifier} has been taken from the work by Zardini et al. \cite{zardini2022implementation}. An example circuit for a toy dataset is shown in \cref{fig:q-cos-class-circ}, and additional details about the implementation can be found in the original article. 

Instead, the quantum distance classifier illustrated in \cref{subsubsec:q-distance-classifier} has been implemented from scratch. An example circuit for the same toy dataset is provided in \cref{fig:q-dist-class-circ}. In detail, the circuit is initialized by computing and directly setting the amplitudes of all qubits except the one ($q_6$) used to encode the training labels. Indeed, the labels are subsequently encoded in the circuit by applying $NOT$ ($X$) and multi-controlled $NOT$ gates. Then, a Hadamard gate ($H$) is applied to the ancillary qubit ($q_0$), and the state of the ancillary and the label qubits is measured. It is worth noting that, at the time of running the experiments, the conditional measurement required by the algorithm was not supported by Qiskit. Hence, all the iterations in which the outcome of the first measurement is $1$ must be discarded when computing the probability estimate. In practice, if the data is standardized, the probability of obtaining $0$ is around $0.5$; otherwise, it is larger than $0.5$. 

Concerning the quantum $k$-NN described in \cref{subsubsec:q-k-nn-classifier}, it has also been implemented from scratch. An example circuit for the same toy dataset is displayed in \cref{fig:q-k-nn-class-circ}. In practice, the circuit is initialized by directly setting the amplitudes of two states: the state encoding the training set ($q_1$-$q_4$), and the state encoding the test instance ($q_5$-$q_6$). After that, a SWAP test is applied to the features registers of the two states ($q_3$-$q_4$ and $q_5$-$q_6$), with the ancillary qubit ($q_0$) as control qubit. Eventually, the state of the ancillary qubit and the state of the index register ($q_1$-$q_2$) are measured.

\section{Empirical Evaluation}
\label{sec:empirical-evaluation}
This section deals with the methods taken into account, the experimental setup used, the datasets considered, and the results obtained.

\subsection{Methods and experimental setup}
\label{subsec:methods-and-exps-setup}
The ensemble techniques and the quantum classifiers considered in this work are summarised in \cref{tab:methods,tab:q-classifiers}, where \textit{quantum\_3NN} represents the quantum $k$-NN model with $k=3$. Instead, \cref{tab:normalizations} lists all the data normalization techniques taken into account. In particular, \textit{none} corresponds to no normalization, \textit{std} stands for standardization, and \textit{minmax} is the so-called min-max normalization. More in detail, the standardization of the $j$-th feature of the $i$-th training instance is defined as
\begin{equation*}
    std(x_{ij}) = \frac{x_{ij} - mean_{t \in train}(x_{tj})}{std_{t \in train}(x_{tj})},
\end{equation*}
whereas the corresponding min-max normalization is given by
\begin{equation*}
    minmax(x_{ij}) = \frac{x_{ij} - \min_{t \in train}{(x_{tj})}}{\max_{t \in train}{(x_{tj})} - \min_{t \in train}{(x_{tj})}}.
\end{equation*}
As a consequence, after the standardization, the features have zero mean and standard deviation equal to one, while, after the min-max normalization, they belong to the $[0, 1]$ interval (the test features are clipped to $0$ or $1$, if they exceed the interval edges).

All ensemble methods have been evaluated with all data normalization techniques. In addition, bootstrap and boosting have been evaluated with all quantum classifiers, also varying the number of internal classifiers ($N$) and the number of training samples per classifier ($S$). Instead, for stacking, the configuration reported in \cref{tab:stacking-config} has been used.

\begin{table}[t!]
    \centering
    \tcaption{Ensemble techniques (a), quantum classifiers (b), and normalization techniques (c) considered.}
    \label{tab:methods}
    \captionsetup{position=top}
    \subfloat[\label{tab:ensembles}]{
        \centering\footnotesize
        \begin{tabular}{c} \hline
            \textbf{Ensemble techniques} \\ \hline
            bootstrap \\
            boosting  \\
            stacking  \\ \hline
        \end{tabular}
    }
    \quad
    \subfloat[\label{tab:q-classifiers}]{
        \centering\footnotesize
        \begin{tabular}{c} \hline
            \textbf{Quantum classifiers} \\ \hline
            quantum\_cosine\\
            quantum\_distance\\
            quantum\_3NN \\ \hline
        \end{tabular}
    }
    \quad
    \subfloat[\label{tab:normalizations}]{
        \centering\footnotesize
        \begin{tabular}{c} \hline
            \textbf{Normalization techniques} \\ \hline
            none   \\
            std    \\
            minmax \\ \hline
        \end{tabular}
    }
\end{table}

\begin{table}[t!]
    \centering
    \tcaption{Stacking configuration tested.}
    \label{tab:stacking-config}
    \footnotesize
    \begin{tabular}{c c} \hline
        \textbf{Classifier} & \textbf{Normalization} \\ \hline
        \multicolumn{2}{c}{Internal}  \\ \hline
        quantum\_cosine     & std     \\
        quantum\_distance   & std     \\
        quantum\_1NN        & minmax  \\
        quantum\_3NN        & minmax  \\ \hline
        \multicolumn{2}{c}{Meta}      \\ \hline
        quantum\_5NN        & none    \\ \hline
    \end{tabular}
\end{table}

Regarding the quantum models, as stated in \cref{subsec:implementation}, the Aer simulator provided by Qiskit has been used for the execution of the algorithms. In particular, the number of measurements, also known as shots, has been set to 8192, which corresponds to the maximum allowed number of shots on real quantum IBM devices. In addition, no noise model has been taken into account in the simulations; therefore, the results represent a best-case scenario. Eventually, it is worth highlighting that two simulation methods have been considered here: \textit{statevector} and \textit{local simulation}. In the former, the results are obtained by processing the final state vector of the circuit; hence, the probability estimates are exact. Instead, in the latter, the behaviour of the real machine is emulated by sampling state counts from the final probability distribution of the circuit.

All results (if not specified differently) have been collected using a Monte Carlo (leave one group out) cross-validation technique \cite{mccv}, with 10 independent runs and a ``80\% training" - ``20\% validation" dataset split. In particular, for each dataset split, the values of the normalization techniques parameters have been computed on the training set samples.

\subsection{Datasets}
\label{subsec:datasets}
In the experiments, 11 datasets taken from the work by Zardini et al. \cite{zardini2022implementation} have been used. These datasets, whose properties are reported in \cref{tab:datasets}, can be downloaded from the GitHub repository associated to the just mentioned article \cite{q-ml-pip}. In particular, the original versions of these datasets come from the UCI Machine learning Repository \cite{Dua:2019}, and most of them have been preprocessed to make them suitable for a binary classification task. It is also worth mentioning that, for the algorithms tested in this work, the considered datasets lead to quantum circuits with sizes of at most 15 qubits, which can be simulated in a reasonable time. 

\begin{table}[hb]
    \centering
    \tcaption{Datasets considered.}
    \label{tab:datasets}
    \footnotesize
    \begin{tabular}{c c c c} \hline
        \textbf{Name} & \textbf{\# samples} & \textbf{\# features} & \textbf{Class balance} \\ \hline
        iris\_setosa\_versicolor       & 100 & 4  & balanced (50/50)        \\
        iris\_setosa\_virginica        & 100 & 4  & balanced (50/50)        \\
        iris\_versicolor\_virginica    & 100 & 4  & balanced (50/50)        \\
        vertebral\_column\_2C          & 310 & 6  & unbalanced (100/210)    \\
        seeds\_1\_2                    & 140 & 7  & balanced (70/70)        \\
        ecoli\_cp\_im                  & 220 & 7  & unbalanced (77/143)     \\
        glasses\_1\_2                  & 80  & 9  & almost balanced (42/38) \\
        breast\_tissue\_adi\_fadmasgla & 71  & 9  & unbalanced (49/22)      \\
        breast\_cancer                 & 80  & 9  & almost balanced (44/36) \\
        accent\_recognition\_uk\_us    & 80  & 12 & unbalanced (63/17)      \\
        leaf\_11\_9                    & 30  & 14 & almost balanced (14/16) \\ \hline
    \end{tabular}
\end{table}

\subsection{Results}
\label{subsec:results}
The results obtained are presented and discussed in the following sections.

\subsubsection{Bootstrap and boosting hyperparameters} 
\label{subsubsec:bootstrap-and-boosting-hyperparams}
Bootstrap and boosting require to set two hyperparameters, namely, the number of internal classifiers $N$ and the number of training samples for each classifier $S$. These parameters have a heavy impact on the performances of the ensembles. Hence, a grid search has been used in order to find the best configuration. In particular, the values taken into account are $N = [5, 10, 30, 50]$ and $S = [6, 8, 10, 20]$.

\begin{figure}[t!]
    \centering
    \includegraphics[width=0.95\linewidth]{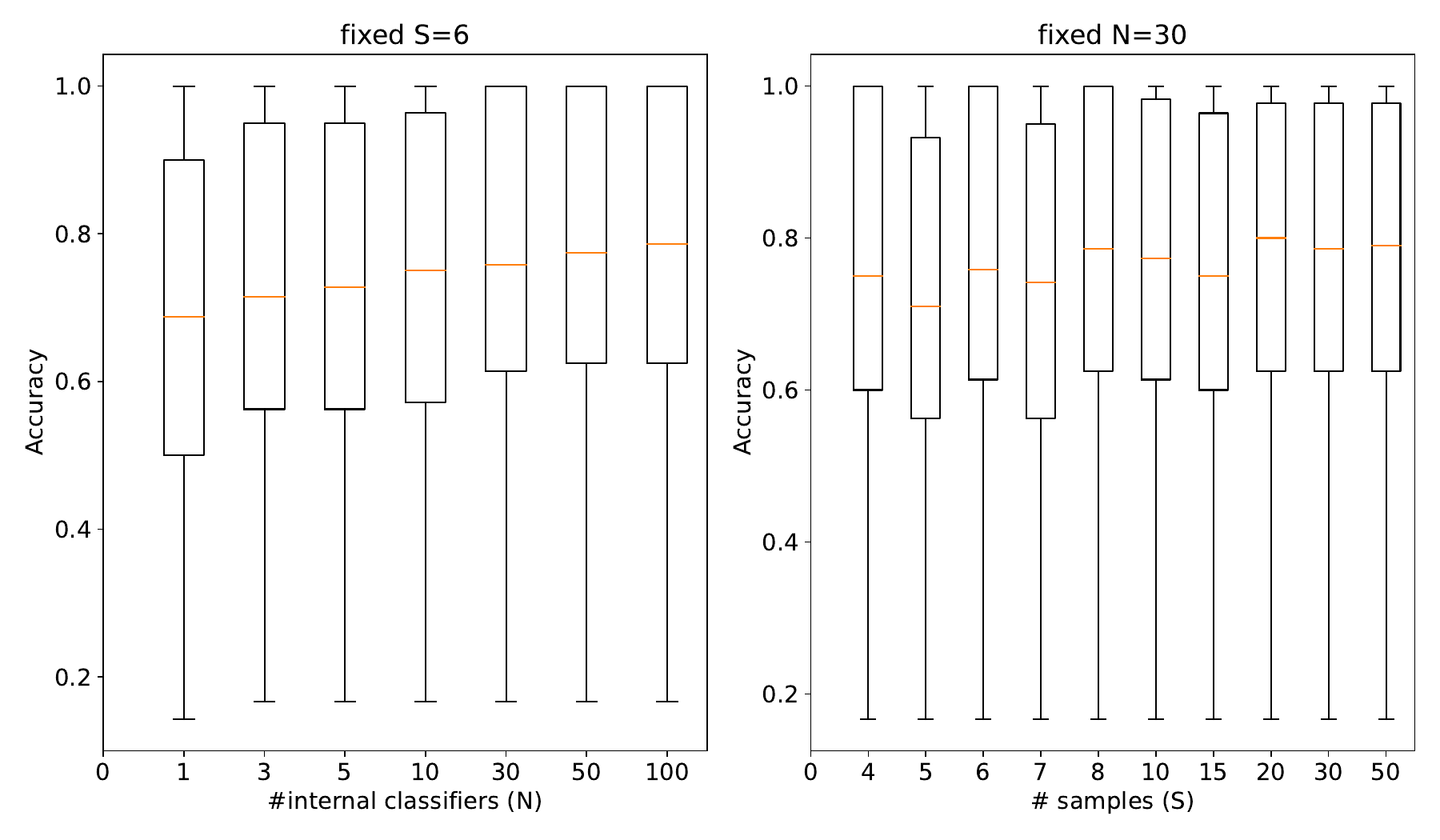}
    \vspace*{5pt}
    \fcaption{Accuracy comparison for the bootstrap technique, varying the number of internal classifiers $N$ (left) and the number of training samples for each classifier $S$ (right) while keeping fixed the other parameter ($S=6$ in the left plot, $N=30$ in the right plot). Each box contains 990 points, with each data point being the accuracy obtained in a run on a certain dataset by a combination of quantum classifier and normalization technique.}
    \label{fig:bootstrap-diff-N-S}
\end{figure}

\begin{figure}[t!]
    \centering
    \includegraphics[width=0.95\linewidth]{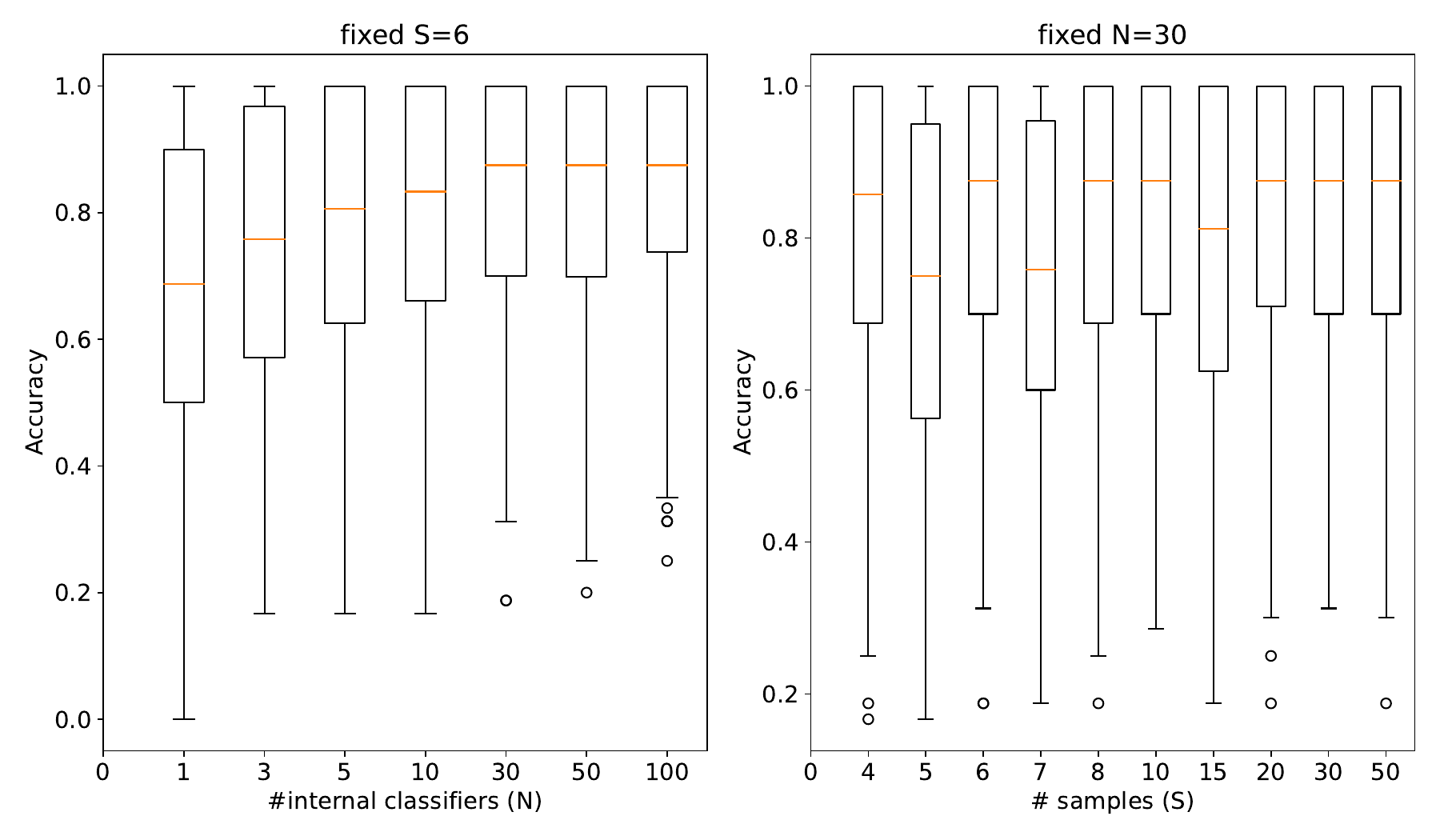}
    \vspace*{5pt}
    \fcaption{Accuracy comparison for the boosting technique, varying the number of internal classifiers $N$ (left) and the number of training samples for each classifier $S$ (right) while keeping fixed the other parameter ($S=6$ in the left plot, $N=30$ in the right plot). Each box contains 990 points, with each data point being the accuracy obtained in a run on a certain dataset by a combination of quantum classifier and normalization technique.}
    \label{fig:boosting-diff-N-S}
\end{figure}

\cref{fig:bootstrap-diff-N-S} (left) shows the accuracy obtained by the bootstrap technique for different $N$ values while keeping fixed the value of $S$ ($S=6$). In particular, each data point represents the accuracy obtained in a run on a certain dataset by a combination of quantum classifier and normalization technique. As expected, the performance improve and the variance decreases by increasing the number of internal classifiers. Instead, for a fixed number of internal models $N$, the performance do not improve by increasing the number of training samples for each classifier, as shown in \cref{fig:bootstrap-diff-N-S} (right). It is also possible to notice that, for even numbers of training samples, the accuracy is almost constant, whereas it drops for odds values. Indeed, odd numbers of training samples imply that the training sets cannot be balanced, since it is a binary classification task. This affects especially the cosine and the distance classifiers, because their prediction is an average value computed over all the training samples; instead, the $k$-NN classifier is less affected by this issue. Moreover, the optimal number of training samples has turned out to be dataset-dependant for the cosine and distance classifiers, while the accuracy of the $k$-NN classifier has always improved by increasing $S$ (as expected).

Analogous plots for the boosting technique are provided in \cref{fig:boosting-diff-N-S}. In detail, the considerations provided for bootstrap about the number of internal classifier and the number of training samples for each classifier hold also for boosting. Actually, by looking at \cref{fig:boosting-diff-N-S} (left), the reduction in variance turns out to be more evident in this case.

In the end, the configuration $N=30,\,S=8$ has been chosen, since $N=30$ already allows achieving good performance and $S=8$ represents a good tradeoff between accuracy and runtime (the size of the index register is three qubits). 

\subsubsection{Performance comparison}
\label{subsubsec:performance-comp}
\begin{figure}[htbp]
    \centering
    \includegraphics[width=0.95\linewidth]{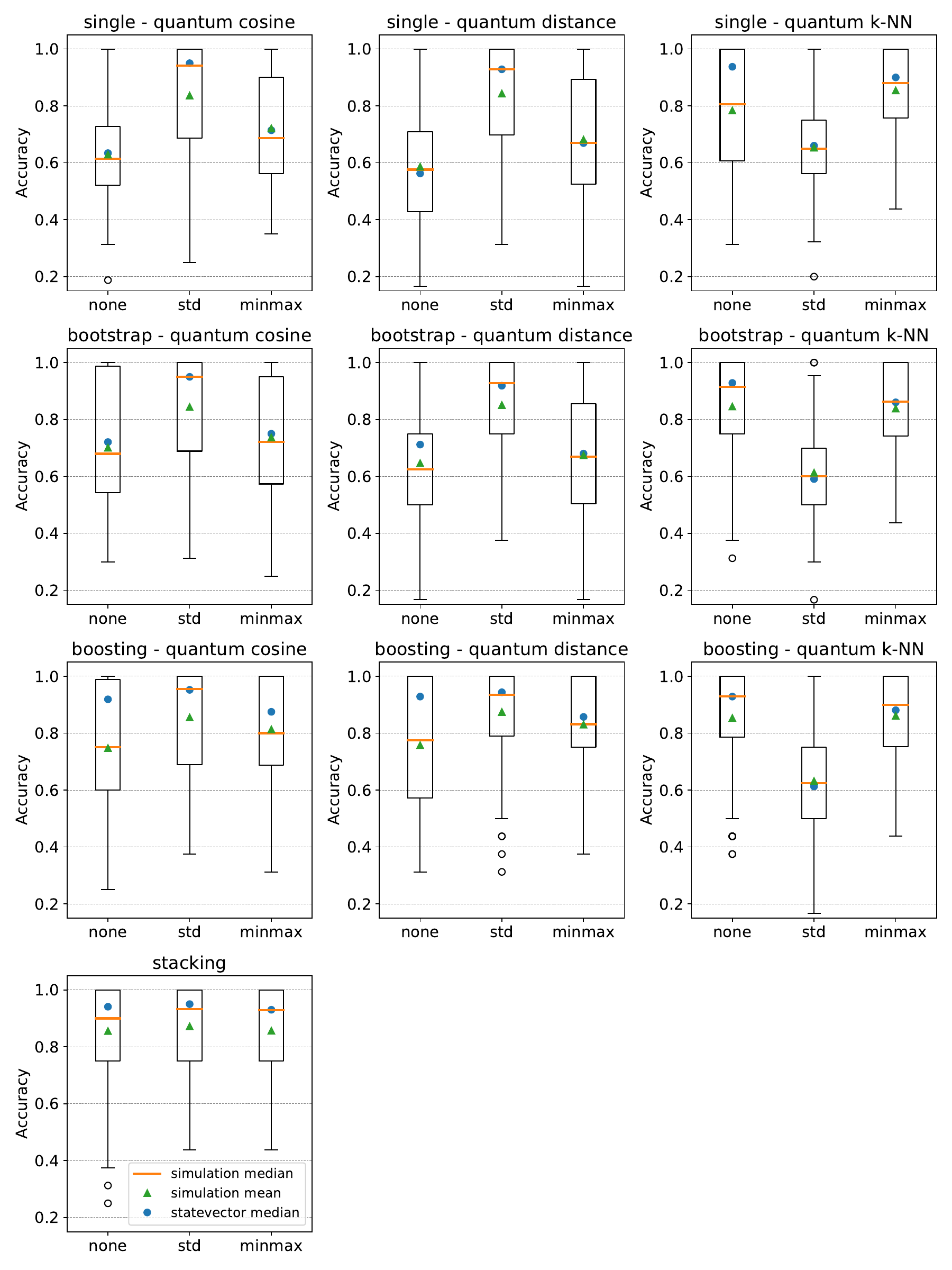}
    \vspace*{10pt}
    \fcaption{Accuracy achieved by \textit{local simulation} with 8192 shots, for all combinations of ensemble, base classifier, and data normalization technique. Each box contains 110 points (one for each run on each dataset).}
    \label{fig:all-perf-boxplots}
\end{figure}

The results achieved by all combinations of ensemble, base classifier, and normalization technique are shown in \cref{fig:all-perf-boxplots}. In detail, each box contains 110 points (one for each run on each dataset), with each data point being the accuracy obtained by \textit{local simulation} with 8192 shots. In addition, the orange line represents the median, the green triangle corresponds to the mean, while the blue circle represents the median of a \textit{statevector} simulation executed on the same data. Focusing on the quantum classifiers executed without ensembles (first row), the quantum cosine and the quantum distance classifiers achieve the best performance when the input data is standardized. Instead, the quantum $k$-NN performs better with a min-max normalization. Indeed, in this quantum $k$-NN, the samples are sorted according to the squared cosine similarity with respect to the test instance; therefore, the features should belong to the same semi-axis to achieve good results.

Concerning bootstrap (second row), the introduction of the ensemble technique leads to a performance improvement for the quantum cosine classifier, while the behaviour with respect to the different normalization techniques remains unchanged. Similar considerations hold for the quantum distance classifier, with the best performance being achieved with the standardization of input data. Instead, the bootstrap ensemble with the quantum $k$-NN performs the best when the input data is not normalized; in this case, there is also a performance improvement with respect to using the single classifier, whereas the performance tend to worsen when a data normalization technique is used. Regarding boosting (third row), the results achieved with the quantum cosine classifier turn out to be better than those obtained by bootstrap and by the single classifier. Indeed, the accuracy is visibly better for no and min-max normalizations, and the variance is lower overall. The best results are still achieved with the standardization of input data. These considerations hold also for the quantum distance classifier, while, for the quantum $k$-NN classifier, boosting and bootstrap turn out to be almost equivalent (there is a little performance improvement). Eventually, the stacking ensemble (fourth row) shows very consistent performance regardless of the data normalization technique employed. Indeed, each internal classifier applies its own normalization technique.

\begin{figure}[htbp]
    \centering
    \includegraphics[width=0.95\linewidth]{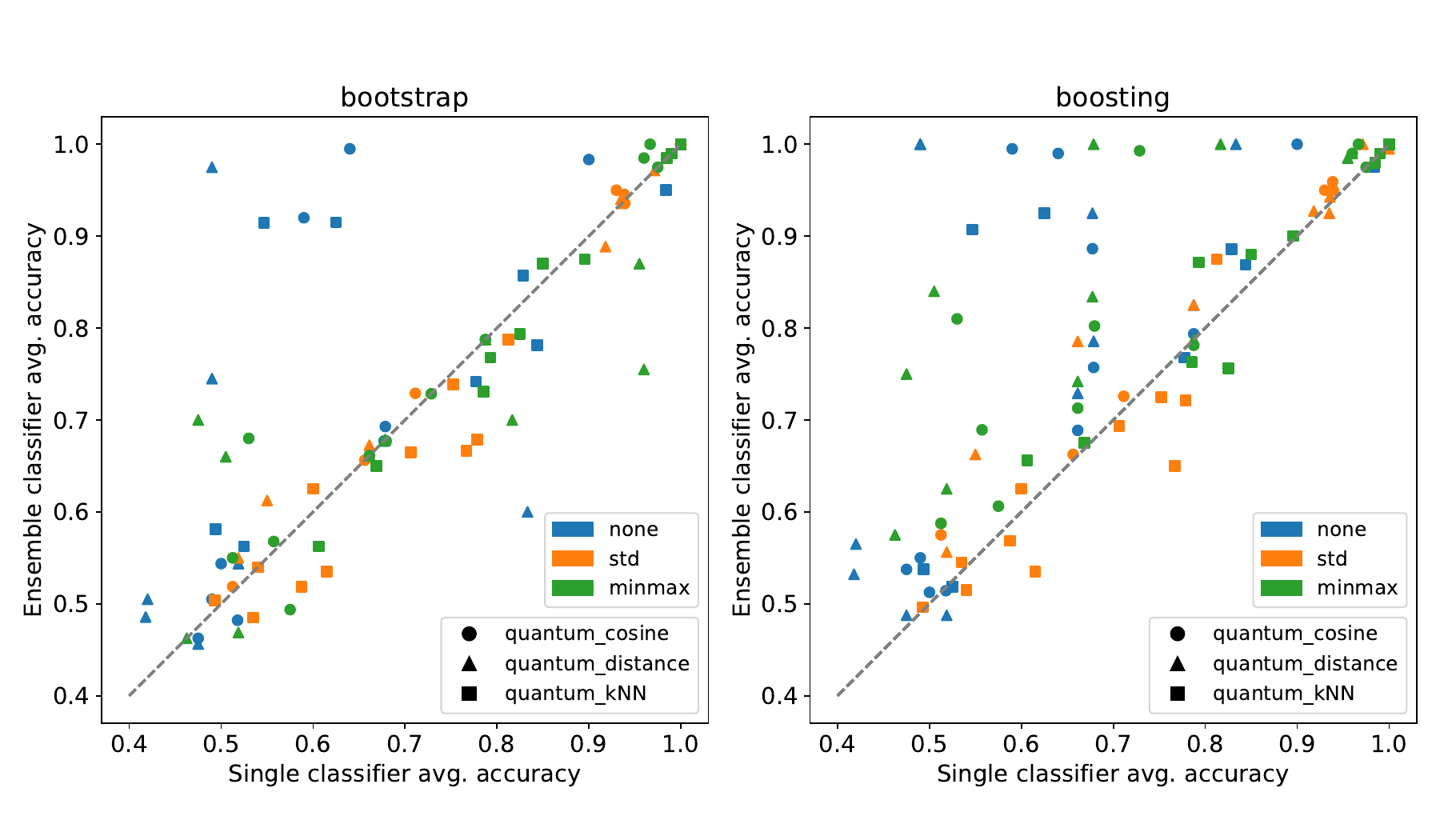}
    \vspace*{10pt}
    \fcaption{Comparison between ensemble techniques and single classifiers in terms of average accuracy over 10 Monte Carlo runs. The left plot refers to bootstrap, the right plot to boosting. In both cases, all normalization techniques are taken into account, each data point corresponds to a different dataset, and \textit{local simulation} with 8192 shots has been used.}
    \label{fig:scatter-advantage}
\end{figure}

\begin{figure}[htbp]
    \centering
    \includegraphics[width=0.95\linewidth]{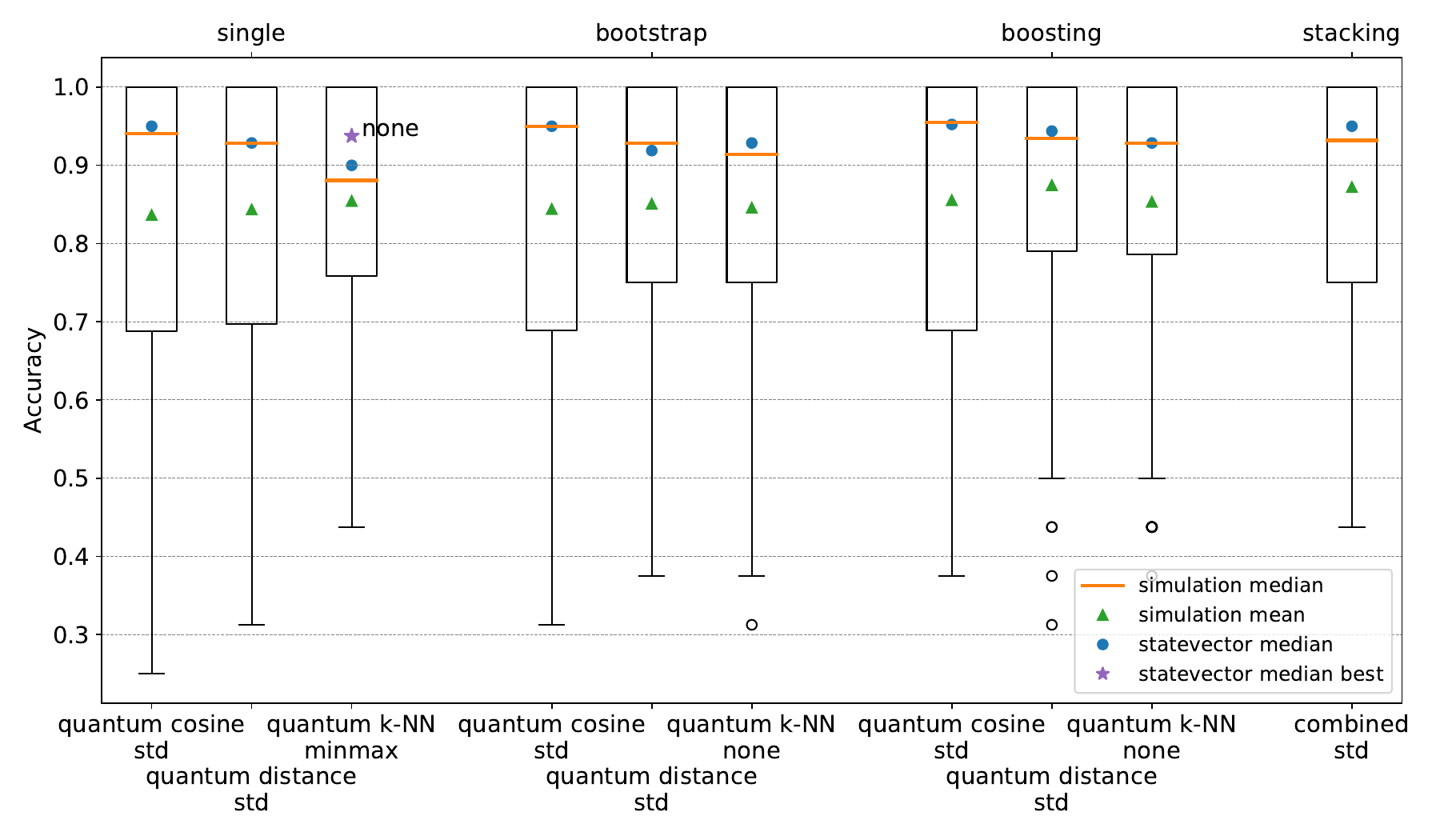}
    \vspace*{10pt}
    \fcaption{Accuracy achieved by \textit{local simulation} with 8192 shots, for all combinations of ensemble and base classifier. Each box contains 110 points (one for each run on each dataset), and only the best-performing normalization technique (in \textit{local simulation}) is shown for each models pair.}
    \label{fig:best-perf-boxplots}
\end{figure}

The effect of introducing bootstrap and boosting with respect to using single quantum classifiers is shown more in detail in \cref{fig:scatter-advantage}. Concerning bootstrap (plot on the left), as stated previously, the performance tend to improve for both the quantum cosine and the quantum distance classifier, whereas they tend to worsen for the quantum $k$-NN. In terms of normalization technique, independently from the classifier used, a significant improvement can be observed for no data normalization. Instead, there is a little advantage for standardization, and an overall neutral effect for min-max normalization. Regarding boosting (plot on the right), the performance of both the quantum cosine and the quantum distance classifiers improve significantly with the introduction of the ensemble technique, whereas there is not a clear benefit for the quantum $k$-NN. As in the previous case, standardization is the only normalization technique not taking advantage of the ensemble usage.

In \cref{fig:best-perf-boxplots}, only the data normalization technique that has achieved the best results in \textit{local simulation} for each ``ensemble" - ``classifier" pair is shown. In this way, it is easier to compare the performance of the different classifiers for each ensemble technique. Focusing on the classifiers without ensembles, the cosine and the distance classifiers with standardization of input data perform in a similar way (the former is a little bit better), and they both outperform the quantum $k$-NN with min-max normalization (the \textit{statevector} median with no data normalization is also reported for the $k$-NN, since it is better than the min-max one). Concerning bootstrap, the classifier achieving the best results (also with respect to the single classifiers) turns out to be the quantum cosine classifier with data standardization. Indeed, both the quantum distance classifier and the quantum $k$-NN show a lower variance, but also a lower median accuracy. Regarding boosting, the configuration with cosine distance classifier and data standardization has the best median accuracy among all methods tested, while the version with distance classifier and data standardization has the best mean accuracy. Eventually, stacking shows good median and mean performance.

\begin{figure}[b!]
    \centering
    \includegraphics[width=0.95\linewidth]{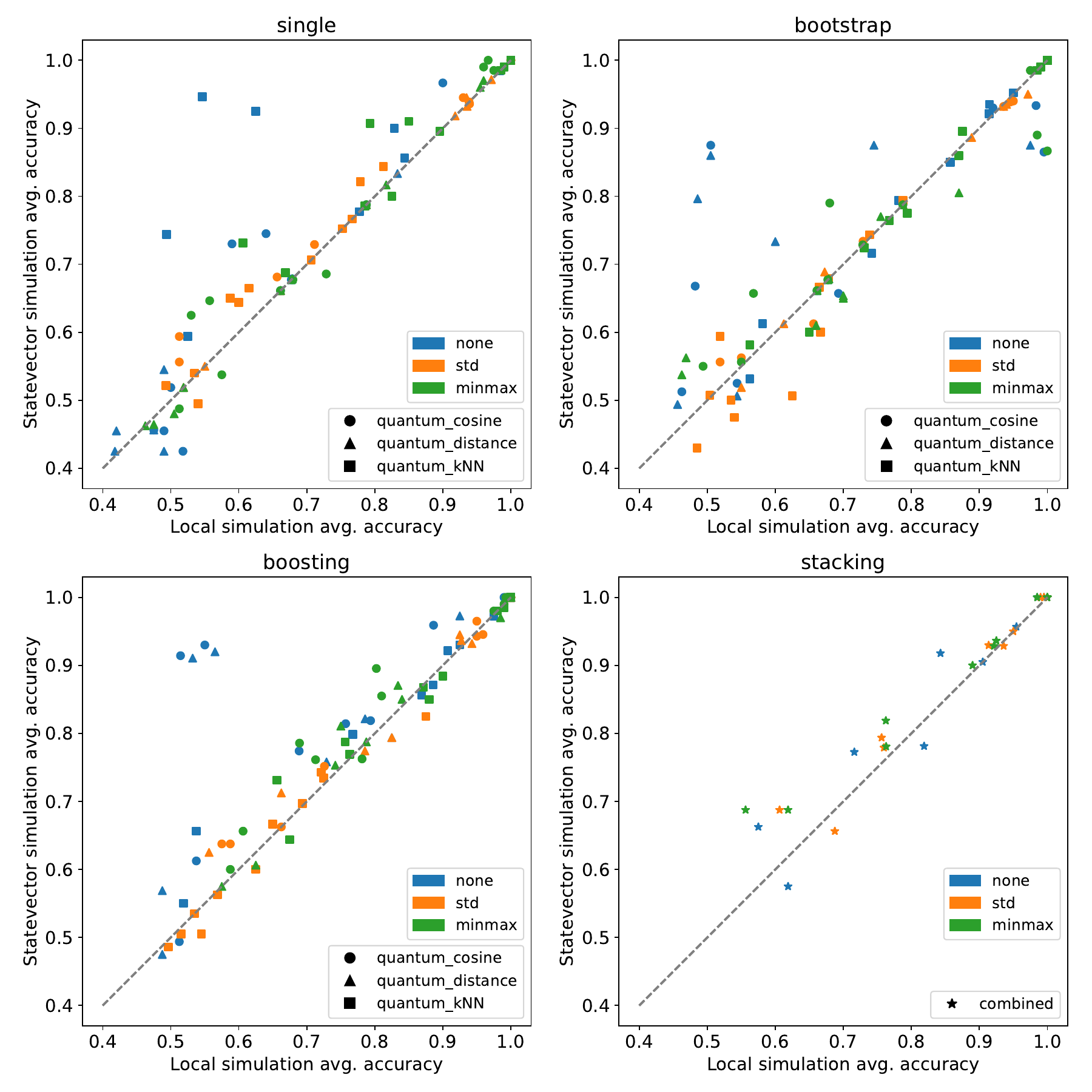}
    \vspace*{10pt}
    \fcaption{Comparison between \textit{statevector} and \textit{local simulation} with 8192 shots in terms of average accuracy over 10 Monte Carlo runs. All combinations of internal classifier and normalization technique are shown for each ensemble technique, and each data point corresponds to a different dataset.}
    \label{fig:scatter-comp}
\end{figure}

\subsubsection{Measurement sampling}
\label{subsubsec:meas-sampling}
As expected, the performance obtained with \textit{local simulation} differ from the ideal ones, represented by \textit{statevector}. Indeed, the repeated sampling from the final probability distribution of the qubits states inevitably introduces some uncertainty, which may lead to a wrong label prediction (for quantum cosine and quantum distance classifiers) or a wrong nearest neighbors ranking (for the quantum $k$-NN). \cref{fig:scatter-comp} shows the impact of sampling on the performance of the various methods. Focusing on the single classifiers (upper left), only in a few cases (located in the low-accuracy region) \textit{local simulation} turns out to be better than \textit{statevector}; in addition, the main outliers are all related to the quantum $k$-NN with no data normalization. Looking at bootstrap (upper right), almost all points are located near the main diagonal, with outliers mainly related to quantum cosine and quantum distance classifiers without data normalization. This suggests that the probability values are close to the decision threshold when the data is not normalized. Concerning boosting (bottom left), the performance drop when using \textit{local simulation} turns out to be evident. This might be related to the ensemble building process; indeed, the uncertainty in the predictions of the internal models might lead to the computation of suboptimal boosting parameters. In addition, the outliers have the same properties of the bootstrap's ones. Eventually, for stacking (bottom right), it is possible to notice a performance drop in the low-middle accuracy region.

\begin{figure}[htbp]
    \centering
    \includegraphics[width=0.95\linewidth]{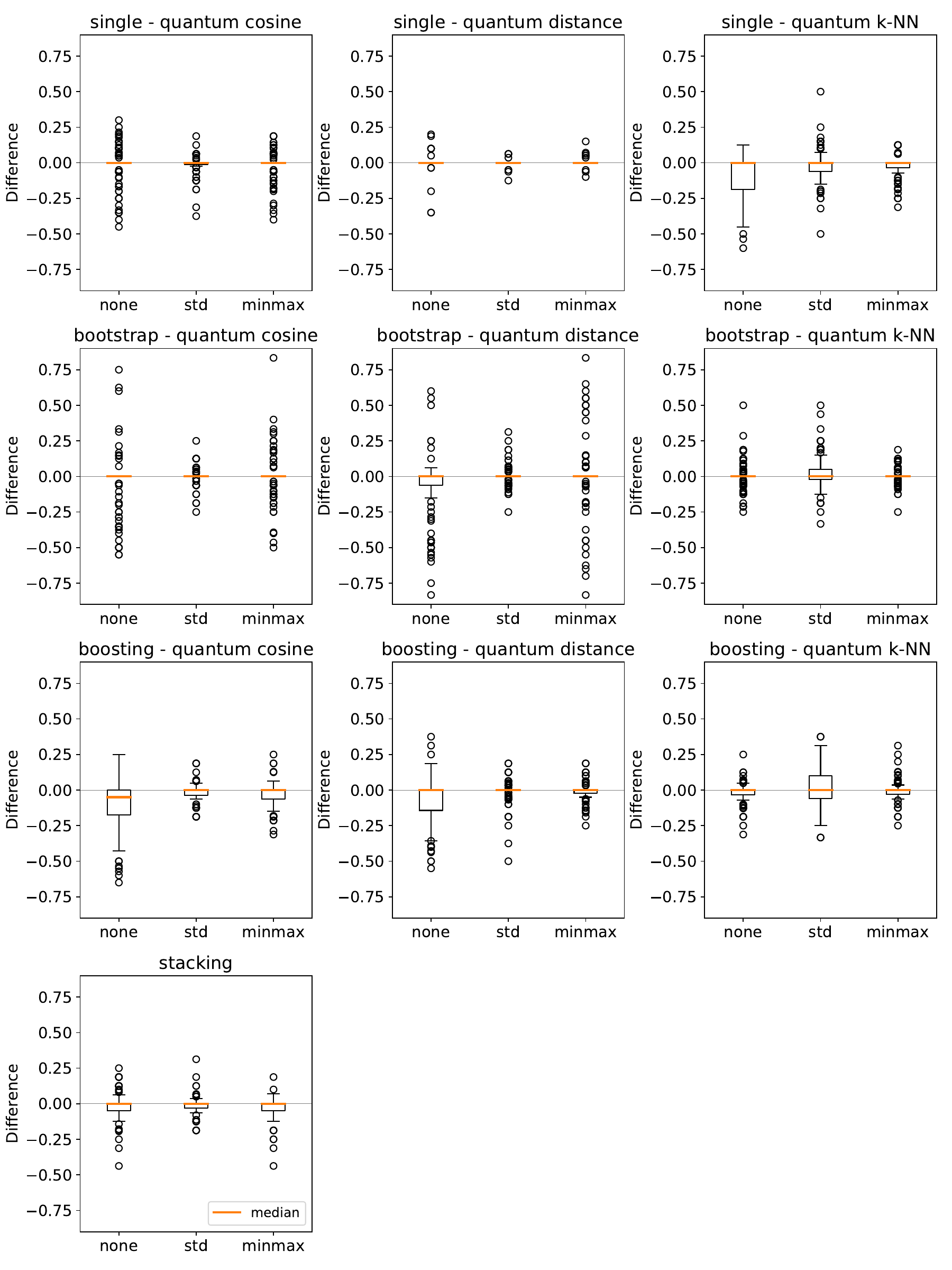}
    \vspace*{10pt}
    \fcaption{Comparison of \textit{local simulation} with 8192 shots and \textit{statevector} in terms of accuracy difference for all combinations of ensemble technique, quantum classifier, and data normalization technique. The number of points per box is 110, with each data point being related to a run on a certain dataset.}
    \label{fig:diff-boxplots}
\end{figure}

To better understand the impact of sampling on each configuration, the distribution of the accuracy difference between \textit{local simulation} and \textit{statevector} for each combination of ensemble technique, quantum classifier, and data normalization is displayed in \cref{fig:diff-boxplots}. The boxplots related to the single classifiers (first row) confirm what has been observed in the previous plots, with the quantum $k$-NN being the only method severely affected by measurement sampling (especially with no data normalization). Regarding bootstrap (second row), it is worth highlighting the performance drop for the quantum distance classifier without data normalization and the accuracy improvement for the quantum $k$-NN with standardization (whose performance in the ideal case are really poor, as shown in \cref{fig:scatter-comp}). In addition, the stability of the quantum $k$-NN with respect to the version without ensemble turns out to be improved. Similar considerations hold also for boosting (third row). Nevertheless, as stated previously, the performance worsening is more marked than for bootstrap. In particular, the quantum cosine classifier turns out to be more affected. Concerning stacking (fourth row), it shows quite good performance in simulation, confirming its stability (especially with standardized data).

\begin{figure}[ht]
    \centering
    \includegraphics[width=0.95\linewidth]{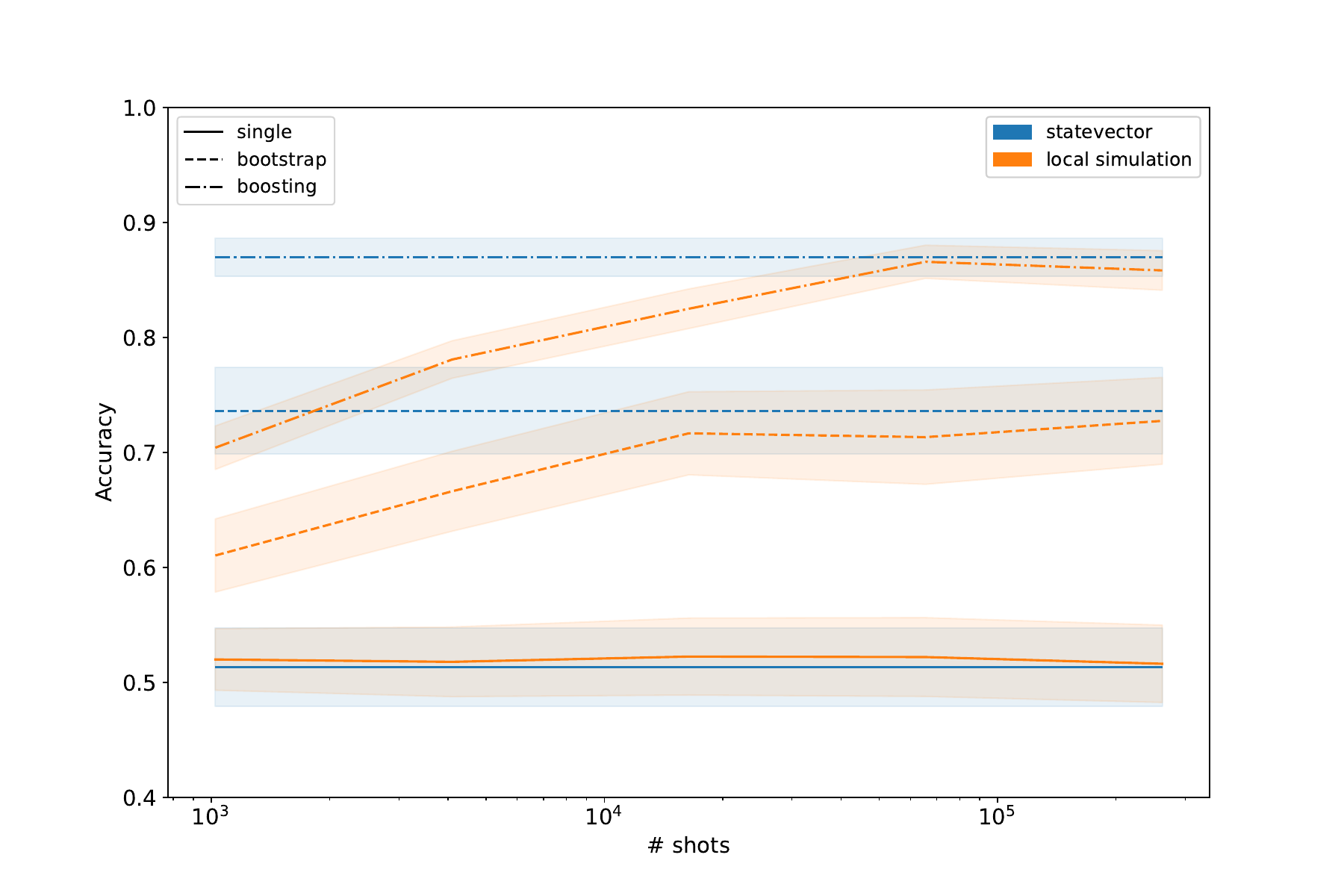}
    \vspace*{10pt}    
    \fcaption{Average accuracy over 120 runs on the \textit{iris\_versicolor\_virginica} dataset as a function of the number of shots (ranging from 1024 to 262144). The plot refers to the quantum distance classifier with min-max data normalization, considering all ensemble techniques; the shaded regions correspond to 95\% confidence intervals, which have been computed according to a Binomial distribution.}
    \label{fig:accuracy-shots-plot}
\end{figure}

Eventually, the relationship between number of measurements (shots) and performance has been analysed for a single case, namely, the quantum distance classifier with min-max data normalization. For this purpose, all ensemble techniques and a single dataset have been taken into account. The results are shown in \cref{fig:accuracy-shots-plot}. In detail, the single classifier achieves comparable results to \textit{statevector} already with $1024$ shots, while, for bootstrap and boosting, the number of measurements required to reach the ideal performance is in the order of $10^5$. Moreover, bootstrap performs worse than boosting, but it reaches the 95\% confidence interval with $10^4$ shots. Instead, boosting achieves comparable performance to bootstrap in the \textit{statevector} execution even with a low number of shots, but it requires ten times repetitions to reach the 95\% confidence interval. In general, a large number of shots is needed to achieve near-to-exact performance (the number of shots grows quadratically with respect to the size of the confidence interval).

\begin{figure}[t!]
    \centering
    \includegraphics[width=0.95\linewidth]{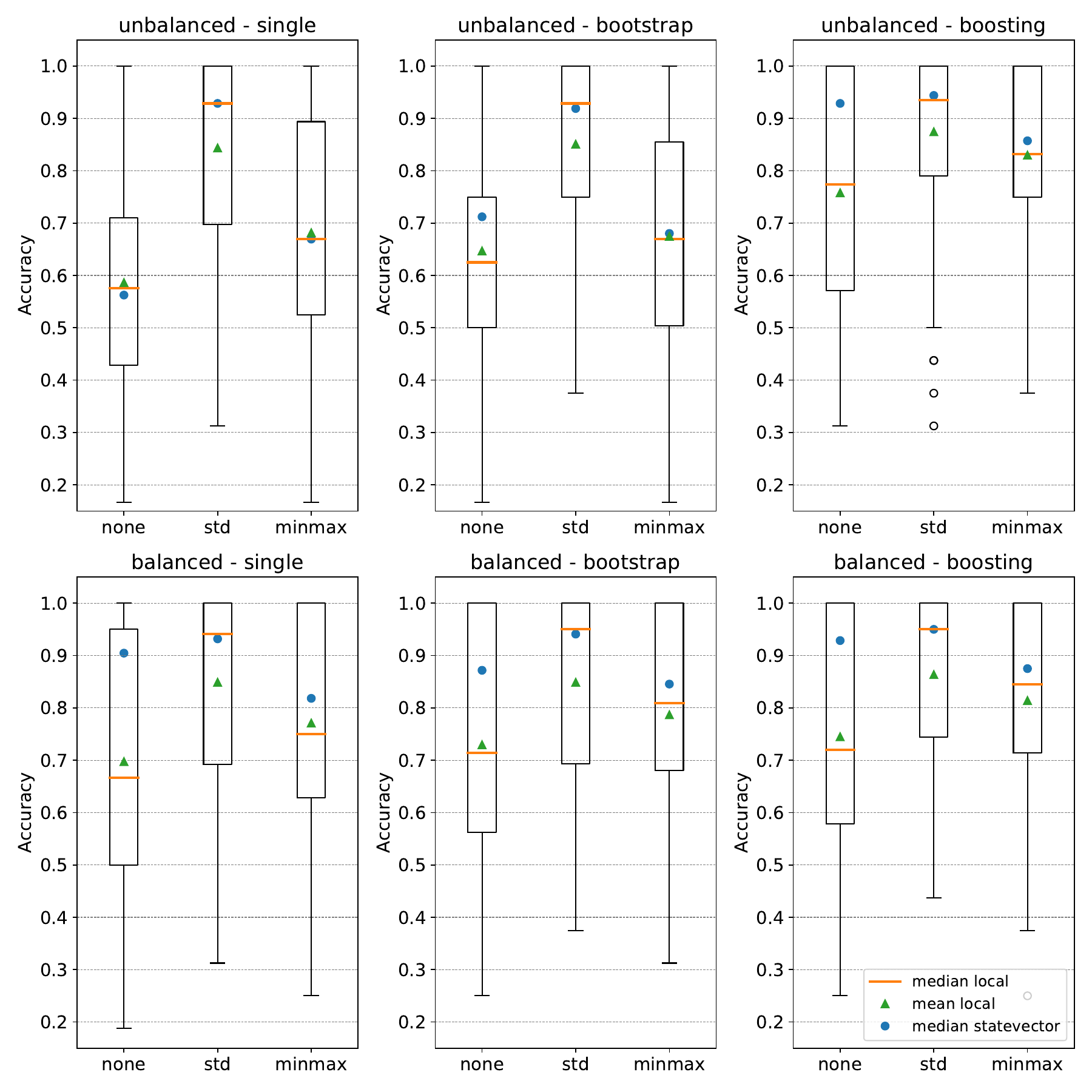}
    \vspace*{10pt}
    \fcaption{Accuracy achieved by the quantum distance classifier with non-balanced and balanced training sets, for all combinations of ensembles and data normalization techniques. The number of shots used for \textit{local simulation} is 8192, and the number of points per box is 110 (one for each run on each dataset).}
    \label{fig:unbalanced-vs-balanced}
\end{figure}

\subsubsection{Class imbalance}
\label{subsubsec:class-imbalance}
Both the quantum cosine and the quantum distance classifiers are strongly influenced by the training data class balance. Indeed, the label predictions are given according to a weighted sum over the training set (Eqs. (\ref{eq:cos-class-classical-formula}) and (\ref{eq:dist-class-classical-formula})). Hence, if the classes are not balanced, the most frequent one is preferred. Moreover, the cosine similarity can take negative values. Instead, in the quantum distance classifier, the weight belongs to the interval $[0, 1]$; as a consequence, every data instance gives a non-negative contribution to the corresponding label. In conclusion, if the weights magnitude is small, the prediction is determined mainly by the ratio of classes in the training set.

\cref{fig:unbalanced-vs-balanced} shows the results achieved by the quantum distance classifier with non-balanced and balanced training sets, for all combinations of ensemble and data normalization techniques. In particular, for the single classifier, the most frequent class has been subsampled in order to match the number of instances of the other class; instead, for the ensembles, the balance has been obtained by sampling the same number of data instances from each class when building the training sets for the internal models. Focusing on the single classifiers, it is possible to notice that, while the results with data standardization are almost identical in the two cases, there is a significant performance improvement for no and min-max data normalizations when forcing the class balance. Nevertheless, a remarkable difference with respect to the ideal performance can be observed. Furthermore, standardization (\textit{std}) turns out to be still the best normalization technique for the quantum distance classifier. Similar considerations hold for bootstrap, whose ideal performance with balanced data are also not always better than the ideal ones for the single classifier; however, the results achieved in simulation are still better. Eventually, boosting seems to be not as sensitive to class imbalance as the other classifiers, and its results are in line with the ones achieved with non-balanced class. In the end, it remains the best-performing ensemble technique.

\section{Conclusions}
\label{sec:conclusions}
In this work, a hybrid ensemble scheme characterised by classical ensemble techniques and quantum classification algorithms has been introduced and empirically evaluated, in simulations without noise, on a binary classification task. In particular, the ensemble techniques taken into account are bootstrap, boosting, and stacking, while the considered quantum classifiers are a quantum cosine classifier, a quantum distance classifier, and a quantum $k$-nearest neighbors classifier. In addition, three data normalization techniques, namely, no normalization, standardization, and min-max normalization, have been taken into account. The results have shown that the introduction of the ensemble techniques leads to a performance improvement with respect to using single quantum classifiers. In detail, the ensemble models have demonstrated a slight advantage in accuracy compared to single classifiers when considering an ideal execution with a suitable data normalization technique. At the same time, they have shown the ability to mitigate both unsuitable data normalizations and measurements uncertainty, making quantum classifiers more stable. Indeed, a strong dependency of the classifiers performance on the data normalization technique used has been observed, as well as the benefit of having class-balanced datasets for classifiers like the quantum distance one.

More in detail, the single quantum cosine and quantum distance classifiers have achieved the best performance with data standardization, while the quantum $k$-NN has performed better with the other two techniques. Both bootstrap and boosting have improved the performance of the single classifiers, especially when the base performance was poor, with the quantum $k$-NN with data standardization being an anomaly in this sense (the performance did not improve). Actually, boosting has proven to be the best ensemble technique in terms of absolute performance but also the most sensitive to measurements uncertainty (due to the iterative structure). It is also worth mentioning that, for both bootstrap and boosting, the number of shots required to reach near-to-exact results has turned out to be quite high. Eventually, stacking has shown good accuracy results and good stability with respect to different data normalization techniques (the internal classifiers perform a subsequent normalization).

Given the promising results obtained, demonstrating how ensembles mitigate sampling and data normalization issues, thereby enhancing performance while constraining circuit size, an interesting possibility for future work could be the development of quantum ensemble techniques. For instance, quantum ensembles could be implemented by means of variational quantum circuits (VQCs) based on Hardware Efficient Ansatzes (HEAs) \cite{Kandala_2017} with parametrized weights induced by the ensembles parameters.

\section*{Acknowledgements}
\noindent
This work was partially supported by project SERICS (PE00000014) under the MUR National Recovery and Resilience Plan funded by the European Union - NextGenerationEU. In addition, E.Z. was supported by by Q@TN, the joint lab between University of Trento, FBK-Fondazione Bruno Kessler, INFN-National Institute for Nuclear Physics and CNR-National Research Council. E.T. was supported by the MUR National Recovery and Resilience Plan (PNRR) M4C1I4.1, funded by the European Union under NextGenerationEU. Views and opinions expressed are however those of the author(s) only and do not necessarily reflect those of the European Union or The European Research Executive Agency. Neither the European Union nor the granting authority can be held responsible for them.

\setcounter{footnote}{0}
\renewcommand{\thefootnote}{\alph{footnote}}


\begin{thebibliography}{000}

\bibitem{Biamonte_2017}
J. Biamonte and P. Wittek and N. Pancotti and P. Rebentrost and N. Wiebe and S. Lloyd (2017), {\it {Quantum machine learning}},
Nature, Springer Science and Business Media {LLC}, Vol. 549, Num. 7671, pp. 195-202,
10.1038/nature23474.

\bibitem{Preskill_2018}
J. Preskill (2018), {\it Quantum Computing in the {NISQ} era and beyond},
Quantum, Verein zur Forderung des Open Access Publizierens in den Quantenwissenschaften, Vol. 2, pp. 79,
10.22331/q-2018-08-06-79.

\bibitem{1688199}
R. Polikar (2006), {\it Ensemble based systems in decision making},
IEEE Circuits and Systems Magazine, Vol. 6, Num. 3, pp. 21-45,
10.1109/MCAS.2006.1688199.

\bibitem{Schuld2018}
M. Schuld and F. Petruccione (2018), {\it Quantum ensembles of quantum classifiers},
Scientific Reports, Vol. 8, Num. 2772,
10.1038/s41598-018-20403-3.

\bibitem{Abbas2020}
A. Abbas and M. Schuld and F. Petruccione (2020), {\it On quantum ensembles of quantum classifiers},
Quantum Machine Intelligence, Vol. 2, Num. 6,
10.1007/s42484-020-00018-6.

\bibitem{araujo2020quantum}
I. C. S. Araujo and A. J. da Silva (2020), {\it Quantum ensemble of trained classifiers},
arXiv, quant-ph,
2007.09293.

\bibitem{macaluso2022quantum}
A. Macaluso and L. Clissa and S. Lodi and C. Sartori (2022), {\it Quantum Ensemble for Classification},
arXiv, cs.LG,
2007.01028.

\bibitem{10.1007/978-3-319-52289-0_9}
D. Windridge and R. Nagarajan (2017), {\it Quantum Bootstrap Aggregation},
Quantum Interaction, Springer International Publishing, pp. 115-121,
978-3-319-52289-0.

\bibitem{qin2022improving}
R. Qin and Z. Liang and J. Cheng and P. Kogge and Y. Shi (2022), {\it Improving Quantum Classifier Performance in NISQ Computers by Voting Strategy from Ensemble Learning},
arXiv, quant-ph,
2210.01656.

\bibitem{zhang2022efficient}
X. Zhang and M. Wang (2022), {\it An efficient combination strategy for hybird quantum ensemble classifier},
arXiv, quant-ph,
2210.06785.

\bibitem{bootstrap}
L. Breiman (1996), {\it Bagging predictors},
Machine Learning, pp. 123–140,
https://doi.org/10.1007/BF00058655.

\bibitem{schapire2013explaining}
R. Schapire (2013), {\it Explaining adaboost},
Empirical inference, Springer, pp. 37-52.

\bibitem{stacking}
D. H. Wolpert (1992), {\it Stacked generalization},
Neural Networks, Vol. 5, Num. 2, pp. 241-259,
ISSN 0893-6080,
https://doi.org/10.1016/S0893-6080(05)80023-1.

\bibitem{Pastorello_2021}
D. Pastorello and E. Blanzieri (2021), {\it A Quantum Binary Classifier based on Cosine Similarity},
2021 {IEEE} International Conference on Quantum Computing and Engineering ({QCE}),
10.1109/qce52317.2021.00086.

\bibitem{Schuld_2017}
M. Schuld and M. Fingerhuth and F. Petruccione (2017), {\it Implementing a distance-based classifier with a quantum interference circuit},Eventually
{EPL} (Europhysics Letters), {IOP} Publishing, Vol. 119, Num. 6, pp. 60002,
10.1209/0295-5075/119/60002.

\bibitem{Afham2020}
A. Afham and A. Basheer and S. K. Goyal (2020), {\it {Quantum $k$-nearest neighbor machine learning algorithm}}, 
arXiv, 
arXiv:2003.09187v1.

\bibitem{Ma2021}
Y. Ma and H. Song and J. Zhang (2021), {\it Quantum Algorithm for K-Nearest Neighbors Classification Based on the Categorical Tensor Network States},
International Journal of Theoretical Physics, Vol. 60, pp. 1164-1174,
10.1007/s10773-021-04742-y.

\bibitem{Giovannetti_2008}
V. Giovannetti and S. Lloyd and L. Maccone (2008), {\it {Quantum Random Access Memory}},
Physical Review Letters, American Physical Society ({APS}), Vol. 100, Num. 16,
10.1103/physrevlett.100.160501.

\bibitem{https://doi.org/10.48550/arxiv.1307.0411}
S. Lloyd and M. Mohseni and P. Rebentrost (2013), {\it {Quantum algorithms for supervised and unsupervised machine learning}},
arXiv,
10.48550/ARXIV.1307.0411.

\bibitem{nielsen_chuang_2010}
M. A. Nielsen and I. L. Chuang (2010), {\it {Quantum Computation and Quantum Information: 10th Anniversary Edition}},
Cambridge University Press,
10.1017/CBO9780511976667.

\bibitem{PhysRevLett.87.167902}
H. Buhrman and R. Cleve and J. Watrous and R. de Wolf (2001), {\it {Quantum Fingerprinting}},
Phys. Rev. Lett., American Physical Society, Vol. 87, Num. 16, pp. 167902-167905,
10.1103/PhysRevLett.87.167902.

\bibitem{1053964}
T. Cover and P. Hart (1967), {\it {Nearest neighbor pattern classification}},
IEEE Transactions on Information Theory, Vol. 13, Num. 1, pp. 21-27,
10.1109/TIT.1967.1053964.

\bibitem{Qiskit}
M. S. Anis and H. Abraham and AduOffei and R. Agarwal and G. Agliardi and others (2021), {\it {Qiskit: An Open-source Framework for Quantum Computing}},
10.5281/zenodo.2573505.

\bibitem{ibmquantum}
IBM (2021), {\it {IBM Quantum}},
https://quantum-computing.ibm.com/.

\bibitem{zardini2022implementation}
E. Zardini and E. Blanzieri and D. Pastorello (2023), {\it Implementation and empirical evaluation of a quantum machine learning pipeline for local classification},
PLOS ONE, Vol. 18, Num. 11, pp. 1-28,
https://doi.org/10.1371/journal.pone.0287869.

\bibitem{mccv}
Q. Xu, Y. Liang (2001), {\it Monte Carlo cross validation},
Chemometrics and Intelligent Laboratory Systems, Vol. 56, Num. 1, pp. 1-11,
ISSN 0169-7439,
https://doi.org/10.1016/S0169-7439(00)00122-2.

\bibitem{q-ml-pip}
E. Zardini (2022), {\it {Implementation and empirical evaluation of a quantum machine learning pipeline for local classification}},
GitHub, GitHub repository,
https://github.com/ZarHenry96/quantum-ml-pipeline.

\bibitem{Dua:2019}
D. Dua and C. Graff (2017), {\it {UCI} Machine Learning Repository},
University of California, Irvine, School of Information and Computer Sciences,
http://archive.ics.uci.edu/ml.

\bibitem{Kandala_2017}
A. Kandala and A. Mezzacapo and K. Temme and M. Takita and M. Brink and J. M. Chow and J. M. Gambetta (2017), {\it {Hardware-efficient variational quantum eigensolver for small molecules and quantum magnets}},
Nature, Springer Science and Business Media {LLC}, Vol. 549, Num. 7671, pp. 242-246,
10.1038/nature23879.

\end{thebibliography}
\end{document}